# Space-time symmetry violation of the fields in quasi-2D ferrite particles with magnetic-dipolar-mode oscillations

E.O. Kamenetskii

Department of Electrical and Computer Engineering,
Ben Gurion University of the Negev, Beer Sheva, 84105, ISRAEL

September 27, 2009

**Abstract**

In magnetic systems with reduced dimensionality, the effects of dipolar interactions allow the existence of long-range ordered phases. Long-range magnetic-dipolar interactions are at the heart of the explanation of many peculiar phenomena observed in nuclear magnetic resonance, ferromagnetic resonance, and Bose-Einstein-condensate structures. In this paper we show that magnetic-dipolar-modes (MDMs) in quasi-2D ferrite disks are characterized by symmetry breaking effects. Our analysis is based on postulates about a physical meaning of the magnetostatic-potential function $\psi(\vec{r},t)$ as a complex scalar wave function, which presumes the long-range phase correlations. An important feature of the MDM oscillations in a ferrite disk concerns the fact that a structure with symmetric parameters and symmetric basic equations goes into eigenstates that are not space-time symmetric. The proper solutions are found based on an analysis of magnetostatic-wave propagation in a helical coordinate system. For a ferrite disk, we show that while a composition of two helical waves may acquire a geometrical phase over-running of $2\pi$ during a period, every separate helical wave has a dynamical phase over-running of $\pi$ and so behaves as a double-valued function. We demonstrate that unique topological structures of the fields in a ferrite disk are intimately related to the symmetry breaking properties of MDM oscillations. The solutions give the MDM power-flow-density vortices with cores at the disk center and azimuthally running waves of magnetization. One can expect that the proposed models of long-range ordered phases and space-time violation properties of magnetic-dipolar interactions can be used in other magnetic structures, different from the ferromagnetic-resonance system with reduced dimensionality.

PACS: 76.50.+g; 68.65.-k; 11.30.Er

# 1. Introduction

Symmetry violation effects are considered as powerful concepts for many developments in magnetism. In particular, unique magnetoelectric (ME) properties – the coupling between the magnetization and polarization vectors – in some magnetic crystals, arise from dynamical symmetry breakings [1 – 3]. There are also the symmetry violation effects closely related to the vortex states in confined magnetic structures. Different characteristic scales in magnetic dynamics – the scales of the spin (exchange interaction) fields, the magnetostatic (dipole-dipole interaction) fields, and the electromagnetic fields – may define different vortex states in confined magnetic structures [4]. In a view of recent studies of the magnetic-dipolar-mode (MDM) oscillations in quasi-2D ferrite disks [5 – 12], dynamical symmetry breaking effects in such structures can be a subject of a special interest. An important feature of the MDM oscillations in a ferrite disk concerns the fact that a structure with symmetric parameters and symmetric basic equations goes into eigenstates that are not space-time symmetric.

The dynamical symmetry breaking effects and vortex solutions in MDM [or magnetostatic (MS)] oscillations are strongly connected with understanding physics of magnetic dipole-dipole interactions in low-dimensional magnetic systems. The dipolar interactions, normally weak enough to be ignored in bulk magnetic materials, play an essential role in stabilizing long-range magnetic order in quasi-2D systems. Such interactions represent a topical problem in a condensed matter theory. The long-range phase correlations among different parts of a confined magnetic structure are at the heart of the explanation of many peculiar phenomena observed in nuclear magnetic resonance (NMR), ferromagnetic resonance (FMR), and Bose-Einstein-condensate (BEC) structures. In spite of the fact that nature of these magnetic structures is very different, certain similarities in physical models of magnetic-dipolar interactions can be found. There can be both classical and quantum models. A number of existing classical models describing the dynamics behavior of the long-range magnetic dipole interaction do not give a comprehensive solution of the problem since cannot explain the reason for the large-scale magnetic ordering and so leaves open the question of long-range phase correlations [13]. To explain such long-range phase correlations, some quantum models have been used. In these models, the role of quantum entanglement in the magnetic-dipolar interactions arises as an important factor. In confined magnetic structures, macroscopic entanglement of a quantum many-spin system [14] can be considered to be caused by the geometric phases [15, 16].

Recent studies of magnetic-dipolar interactions in a quasi-2D ferrite disk revealed unique properties of eigenmode oscillations. The MDMs are characterized by energy eigenstates [5, 6], gauge electric fluxes and eigen electric (anapole) moments [7]. Special vortex characteristics of MDMs in thin-film ferrite disks were found numerically and analytically [8, 9]. The obtained results give a deep insight into an explanation of the experimental multiresonance absorption spectra shown both in well known previous [17, 18] and new [10 – 12] studies. In a view of these unique spectral properties, a detailed analysis of the dynamical symmetry breaking effects in MDM oscillations appears as a very important subject. In this paper, we show that the MDM oscillations in a ferrite disk are macroscopically entangled states associated with geometric phases. To come to such a conclusion, we make analytical studies of the MDM spectra in cylindrical and helical coordinate systems. We demonstrate that unique topological structures of the fields in a ferrite disk are intimately related to the symmetry breaking properties of the MDM oscillations.

The paper is organized as follows. The paper begins with Section 2 giving an analysis of the known models for long-range magnetic-dipolar interactions in confined magnetic structures. Section 3 is devoted to spectral characteristics of magnetic-dipolar modes in a normally magnetized quasi-2D ferrite disk analyzed in cylindrical and helical coordinate systems. Based on such spectral characteristics, unique topological properties of the MDM fields are shown in Section 4. In Section 5 we discuss the symmetry breaking effects of the MDM oscillations in a ferrite disk. The paper will be concluded by a summary in Section 6.

## 2. The models of long-range magnetic-dipolar interactions in confined magnetic structures

### 2.1 Classical models

It is clear that classically, distant magnetic dipolar fields can be modeled as a sum of the fields produced by the magnetic dipoles. Let us consider localized classical magnetic dipoles. Such a dipole placed in point $j$ and having a magnetic moment $\vec{M}_j$ creates in point $j'$ a magnetic field

$$\vec{H} = \frac{3(\vec{M}_j \cdot \vec{r}_{jj'})\vec{r}_{jj'}}{\vec{r}_{jj'}^5} - \frac{\vec{M}_j}{\vec{r}_{jj'}^3}, \tag{1}$$

where $\vec{r}_{jj'}$ is a radius-vector connecting points $j$ and $j'$. If in a point $j'$, another dipole having a magnetic moment $\vec{M}_{j'}$ is placed, it is affected by a torque

$$\vec{T} = \vec{M}_{j'} \times \vec{H}. \tag{2}$$

Interaction energy for these two magnetic dipoles is

$$\varepsilon_{jj'} = -\vec{M}_{j'} \cdot \vec{H} = -\frac{3(\vec{M}_j \cdot \vec{r}_{jj'})(\vec{M}_{j'} \cdot \vec{r}_{jj'})}{\vec{r}_{jj'}^5} - \frac{\vec{M}_j \cdot \vec{M}_{j'}}{\vec{r}_{jj'}^3}. \tag{3}$$

One can find energy of magnetic interactions for all the magnetic moments in a sample as

$$W_M = \frac{1}{2} \sum_j \sum_{j' \neq j} \varepsilon_{jj'}. \tag{4}$$

This classical interpretation of interaction between magnetic dipoles cannot explain, however, the reason for magnetic ordering [13] and so leaves open the question of *long-range phase correlations* for magnetic-dipolar interactions in a confined magnetic structure.

A classical model described by Eqs. (1) – (4) can be analyzed numerically. The simplest possible numerical model of dipolar interactions involves a discrete lattice of classical spins and direct summation of their magnetic fields. In Ref. [19], authors used the discrete dipole approximation – a numerical approach based on a model involving *N* discrete classical magnetic dipoles. The approach – viewed as a discrete version of the Landau-Lifshitz equation – involves a solution of the coupled

Larmor equations of the individual magnetic dipoles with all the fields (dipole, exchange, and magnetic-anisotropy) acting on them. In this approximation, the dipole fields are considered as "pure" magnetostatic fields. The problem is reduced to solving a system of linear equations. One obtains the eigenfrequencies of the resonant modes and the eigenvectors are the relative amplitudes of excitation of the individual dipoles. Similar micromagnetic strategies are used in Refs. [20], where the solutions are obtained for the spatially nonhomogeneous dipolar field in nonellipsoidal magnetic samples. In the models, used both in Refs. [19] and [20], the concept of phenomenological magnetization is introduced. The boundary conditions for the dipolar-mode fields are imposed on the magnetization and it is supposed that there exist the *long-range magnetization wave propagation*. One has to note, however, that in neglect of the exchange interaction (when the system does not show the space dispersion properties), the Landau-Lifshitz equation contains only the time derivative of magnetization and so there is evident inconsistency in solutions of the dipolar-mode spectral problem obtained with use of the boundary conditions for magnetization. Certainly, from a spectral theory it is known that in a formulation of the boundary-value problem, the form of homogeneous boundary conditions should be in correspondence with to the form of a self-conjugate differential operator (see e. g. [21]). It is also well known that the amplitude of the dynamic magnetization is not fully specified at the boundaries of a magnetic system by usual electrodynamic boundary conditions [22].

Another classical approach in analyzing a long-range mechanism of the dipole-dipole interaction is based on consideration of a magnetic medium as a continuum. Contrary to an exchange spin wave, in magnetic-dipolar waves the local fluctuation of magnetization does not propagate due to interaction between the neighboring spins. When field differences across the sample become comparable to the bulk demagnetizing fields the local-oscillator approximation is no longer valid, and indeed under certain circumstances, entirely new spin dynamics behavior in a magnetic medium can be observed. This dynamics behavior in a magnetic sample is the following. Precession of magnetization about a vector of a bias magnetic field produces a small oscillating magnetization $\vec{m}$ and a resulting dynamic demagnetizing field $\vec{H}$, which reacts back on the precession, raising the resonant frequency. In the continuum approximation, vectors $\vec{H}$ and $\vec{m}$ are coupled by the differential relation:

$$\vec{\nabla} \cdot \vec{H} = -4\pi \vec{\nabla} \cdot \vec{m}, \qquad (5)$$

where

$$\vec{H} = -\nabla \psi \qquad (6)$$

and $\psi$ is a MS potential. This, together with the Landau-Lifshitz equation, leads to complicated integro-differential equations for the mode solutions in a lossless magnetic sample.

For calculation, the formulation based on the MS-Green-function integral problem for magnetization $\vec{m}$ was suggested and used in Refs. [23 – 25]. In this case one solves "pure static" MS equations for a dipolar field. It is supposed that the sources of a MS field are both volume "magnetic charges", arising from $\nabla \cdot \vec{m}$, and surface "magnetic charges", arising from discontinuity of the normal component of $\vec{m}$ on the surface of a ferrite sample. The MS potential $\psi$ is defined based on integration of the MS Poisson equation

$$\nabla^2 \psi = -4\pi \rho_m, \tag{7}$$

where

$$\rho_m \equiv -\vec{\nabla} \cdot \vec{m}. \tag{8}$$

For the edge and volume sources one obtains [22]:

$$\psi = -\int_V \frac{\vec{\nabla}_{r'} \cdot \vec{m}(\vec{r}\,')}{|\vec{r}-\vec{r}\,'|} dV + \oint_S \frac{\vec{n}' \cdot \vec{m}(\vec{r}\,')}{|\vec{r}-\vec{r}\,'|} dS, \tag{9}$$

where $\vec{n}'$ is the outwardly directed normal.

Such a theoretical analysis of the RF magnetization eigenvalue problem encounters a significant difficulty due to the absence of exact information of the boundary conditions for RF magnetization $\vec{m}$ since, as we discussed above, in classical electrodynamics the boundary conditions are imposed on the normal components of the magnetic induction and the tangential component of the magnetic fields, but not on the components of magnetization. So the dynamic magnetization at the boundary of a magnetic element is undefined from classical electrodynamics. To solve a problem, approximate boundary conditions were derived recently by Guslienko and Slavin [26]. These effective boundary conditions, applicable within a macroscopic approach, are generalized from the exchange boundary conditions of Rado and Weertmann [27]. They are useful when the regions of spatial inhomogeneity in the magnetization are confined near the boundaries of an otherwise uniformly magnetized sample. It becomes evident that the MS spectral problem for long-range dipolar-mode fields, being analyzed based on such boundary conditions, is not the self-conjugate problem. In frames of the above continuum-medium classical approach, the question of long-range phase correlations for magnetic-dipolar interactions remains still open.

The question on the magnetic-dipolar long-range phase correlations arises not only in ferromagnetic bodies. It appears also in the nuclear-magnetic-resonance (NMR) confined structures. In spite of the fact that nature of NMR and FMR structures is very different, the physical models of magnetic-dipolar interactions in such magnetic systems with reduced dimensionality, allowing the existence of long-range ordered phases, can be quite similar. It was pointed out that classical dipolar fields can play a prominent role in NMR of highly magnetized samples at low temperatures. The simplest numerical model of dipolar interactions in NMR structures involves a discrete lattice of spins and direct summation of their magnetic fields [28]. A combined effect of spatial inhomogeneities in the dipolar field with corresponding inhomogeneities in magnetization shows the existence of discrete NMR spectra (so called spectral clustering). The analysis predicts the existence of a set of MS modes which depend upon sample shape and field gradients [29].

From the above analyzed FMR and NMR problems, one can conclude that for a confined magnetic structure with homogeneous material parameters, the MDMs will appear only if (a) exchange interaction takes place and/or (b) an internal DC magnetic field is non-homogeneous. Nevertheless, in 1956, White and Solt showed experimentally that in "big" (when the exchange fluctuations are neglected) spherical ferrite samples, having homogeneous internal DC magnetic field, the MDMs occur [30]. Walker obtained analytical solutions for such MDMs in ferrite spheroids [31]. To solve the problem, Walker used the continuum approximation based on the known [from the linearized local (non-exchange-interaction) Landau-Lifshitz equation] permeability tensor $\vec{\vec{\mu}}$ [13]. The egenvalue problem was formulated based on the differential-operator equation for a *"fictitious"* MS-potential *wave* function $\psi$ ($\vec{H} = -\nabla \psi$):

$$\vec{\nabla} \cdot (\vec{\vec{\mu}} \cdot \vec{\nabla} \psi) = 0. \tag{10}$$

The boundary conditions were imposed on the MS-potential field and not on the RF magnetization. In such a spectral problem, there are MS-potential propagating fields which cause and govern propagation of magnetization fluctuations. In other words, space-time magnetization fluctuations are corollaries of the propagating MS-potential fields, but there is no the magnetization-wave spectral problem. Similar MS-wave approach was used in studies of absorption spectra in NMR samples [32]. An analysis of the MDM boundary-value problem based on the second-order differential equation (10) does not have evident contradictions since the boundary conditions for the MS-potential wave function are in correspondence with to the form of a differential operator. Nevertheless, the physical meaning of the MS-potential wave function $\psi(\vec{r},t)$ (which presumes the fact of the long-range phase correlations) in confined magnetic structures is unclear from classical electrodynamics models.

## 2.2. Long-range phase correlation and macroscopic quantum entanglement

Generally, in classical electromagnetic problems for time-varying fields, there are no differences between the methods of solutions: based on the field representation or based on the potential representation of the Maxwell equations. For the wave processes, in the field representation we solve a system of first-order partial differential equations (for the electric and magnetic vector fields), while in the potential representation we have a smaller number of second-order differential equations (for the scalar electric or vector magnetic potentials). The potentials are introduced as formal quantities for a more convenient way to solve the problem and a set of equations for potentials are equivalent in all respects to the Maxwell equations for fields [22]. The situation becomes completely different, however, if one supposes to solve a boundary-value problem for electromagnetic wave processes in small samples of a strongly temporally dispersive magnetic medium [33]. In this case one has MS waves. There are the magnetic-dipole oscillations [13]. The fact that in the MS-wave processes one has negligibly small variation of the electric energy [33] raises the questions about the nature of the RF electromagnetic fields in small ferrite samples. In particular, the physical meaning of the electromagnetic power flow (the Poynting vector) becomes unclear since for MS-wave processes there are no real mechanisms of transformation of the curl electric field to the MS-potential magnetic field. For magnetic samples with the MS resonance behaviors in microwaves, the spectral problem cannot be formally reduced to the complete-set Maxwell-equation representation. A "fictitious" MS-potential wave function $\psi$, which describes MS waves in small ferrite samples, does not have a proper justification in classical electromagnetic problems.

The fact that proper classical interpretation of interaction between magnetic dipoles in a confined magnetic structure cannot explain the reason for magnetic ordering, leaves open the question of long-range phase correlations described by MS-potential wave function $\psi(\vec{r},t)$. At the same time, such a wave function is not properly related to exchange-interaction spin waves. It is well known that magnetostatic ferromagnetism has a character essentially different from exchange ferromagnetism [13, 20, 34]. The MDM spectral properties shown in Refs. [5 – 7] are based on postulates about a physical meaning of MS-potential functions $\psi(\vec{r},t)$ as complex scalar wave functions with energy-eigenstate orthogonality conditions. The model [5 – 7], in which a static internal magnetic field is regarded as a *potential well* for MS-potential wave functions, gives an important insight into the essence of the MDM localization. With the Hilbert-space spectral properties of scalar wave functions $\psi(\vec{r},t)$, one can explain experimentally observed discrete energy states and eigen electric moments of oscillating MDMs [10 – 12, 17, 18].

From a quantum point of view, it is evident that a ferrite sample is a many-spin system with strong quantum fluctuations. An important goal is to understand the dynamics of such a many-body macroscopic system. In a view of discussed difficulties in classical explanations of long-range phase correlations for dipole-dipole interactions in confined magnetic structures, the problem founds unexpected solution in recent studies of macroscopic quantum entanglements in a many-spin system. Entanglement is a striking feature of quantum mechanics revealing the existence of nonlocal correlations among different parts of a quantum system. It appears that when long-range phase correlations for dipole-dipole interactions in a quasi-2D ferrite disk take place, one has certain entangle states for precessing spins. The entangled many-body states of a "big" (when sizes of a sample are much larger than characteristic sizes of the exchange-interaction properties) confined magnetic system may suggest new types of magnetic-dipolar collective phenomena. When an intense homogeneous DC magnetic field aligns all the spins of a ferromagnet along its direction, a ferromagnetic sample is characterized by an unentangled (or product) state. When, however, a spin-chain system with a long-range dipole-dipole interaction is placed into a nonhomogeneous DC magnetic field, the entanglement can occur [35]. Another situation of spin entanglement can be caused by the geometric phases. It was shown that the concurrence for the entanglement of two distinguishable spins can be formulated in terms of the Berry phase acquired by the spins when each spin is rotated about the quantization axis [15, 16, 36].

For our studies of long-range phase correlations, the most interesting model is a model of spin entanglement due to geometric phases. In a quasi-2D ferrite disk, the MDMs are characterized by discrete energies of macroscopically distinct states. Entanglement of interacting spins is strongly correlated with macroscopic properties of these distinct states and the many-spin systems are regarded as systems which are arranged on topological structures. The fact of macroscopic quantum entanglement for long-range magnetic-dipolar interactions in a quasi-2D ferrite disk becomes evident from an analysis of the MDM spectral problems.

## 3. Magnetic-dipolar modes in a normally magnetized quasi-2D ferrite disk

A phenomenological theory can provide an important guidance in understanding the macroscopic properties of a many-body system. Our phenomenological theory is based on an analysis of spectral properties of MS-potential wave functions in quasi-2D ferrite disks. In such an analysis, one becomes faced, however, with nonintegrability (path dependence) of the problem. Such nonintegrability appears because of a special phase factor on a lateral border of a normally magnetized ferrite disk.

In an initial formal assumption of separation of variables for MS-potential wave functions in a quasi-2D ferrite disk, a spectral problem in cylindrical coordinates $z, r, \theta$ is formulated with respect to membrane MS functions (described by coordinates $r, \theta$) with amplitudes dependable on $z$ coordinate. For a dimensionless membrane MS-potential wave function $\tilde{\varphi}$, the boundary condition of continuity of a radial component of the magnetic flux density on a lateral surface of a ferrite disk of radius $\Re$ is expressed as [6, 7]:

$$\mu\left(\frac{\partial \tilde{\varphi}}{\partial r}\right)_{r=\Re^-} - \left(\frac{\partial \tilde{\varphi}}{\partial r}\right)_{r=\Re^+} = -i\frac{\mu_a}{\Re}\left(\frac{\partial \tilde{\varphi}}{\partial \theta}\right)_{r=\Re^-}, \qquad (11)$$

where $\mu$ and $\mu_a$ are, respectively, diagonal and off-diagonal components of the permeability tensor $\ddot{\mu}$. The term in the right-hand side (RHS) of Eq. (11) has the off-diagonal component of the permeability tensor, $\mu_a$, in the first degree. There is also the first-order derivative of function $\tilde{\varphi}$ with respect to the azimuth coordinate. It means that for the MS-potential wave solutions one can distinguish the time direction (given by the direction of the magnetization precession and correlated with a sign of $\mu_a$) and the azimuth rotation direction (given by a sign of $\partial \tilde{\varphi}/\partial \theta$). For a given sign of a parameter $\mu_a$, there are different MS-potential wave functions, $\tilde{\varphi}^{(+)}$ and $\tilde{\varphi}^{(-)}$, corresponding to the positive and negative directions of the phase variations with respect to a given direction of azimuth coordinates, when $0 \leq \theta \leq 2\pi$. Let, for a given direction of a bias magnetic field, a certain azimuthally running magnetostatic wave acquires a phase $\Phi_1$ after rotation around a disk. For an opposite direction of a bias magnetic field such a phase will be $\Phi_2$. It is evident that $|\Phi_1| = |\Phi_2| \equiv \Phi$ and there should be $\Phi_1 + \Phi_2 = 2\pi n$ or $\Phi = n\pi$. Quantities $n$ are odd integers. This follows from the time-reversal symmetry breaking effect. A system comes back to its initial state after a full $2\pi$ rotation. But this $2\pi$ rotation can be reached if both partial rotating processes, with phases $\Phi_1$ and $\Phi_2$, are involved. So minimal $n=1$ and, generally, quantities $n$ are odd integers. From the above consideration, one may come to conclusion that for a given direction of a bias magnetic field, a membrane function $\tilde{\varphi}$ behaves as a double-valued function.

To make the MS-potential wave functions single-valued and so to make the MDM spectral problem analytically integrable, two approaches were suggested. These approaches, distinguishing by differential operators and boundary conditions used for solving the spectral problem, give two types of the MDM oscillation spectra in a quasi-2D ferrite disk.

### 3.1. *G*- and *L*-mode magnetic-dipolar oscillations in a ferrite disk: A cylindrical coordinate system

In frames of the first approach, we describe the spectral problem by the second-order differential-operator equation [5 – 7]

$$\left(\hat{G}_\perp - (\beta)^2\right)\tilde{\eta} = 0, \qquad (12)$$

where

$$\hat{G}_\perp \equiv \mu \nabla_\perp^2, \qquad (13)$$

$\nabla_\perp^2$ is the two-dimensional (with respect to in-plane coordinates) Laplace operator, $\tilde{\eta}(r,\theta)$ is a dimensionless membrane MS-potential wave function within the domain of definition of operator $\hat{G}_\perp$, and $\beta$ is a propagation constant along $z$ axis ($\eta \equiv \tilde{\eta}\, e^{-i\beta z}$). Operator $\hat{G}_\perp$ is positive definite for negative quantities $\mu$. Outside a ferrite region Eq. (12) becomes the Laplace equation ($\mu = 1$).

Double integration by parts on square $S$ – an in-plane cross-section of a disk structure – of the integral $\int_S (\hat{G}_\perp \tilde{\eta}) \tilde{\eta}^* dS$ gives the boundary conditions for self-adjointess of operator $\hat{G}_\perp$. The corresponding boundary conditions for a disk of a radius $\Re$ are:

$$\mu \left( \frac{\partial \tilde{\eta}}{\partial r} \right)_{r=\Re^-} - \left( \frac{\partial \tilde{\eta}}{\partial r} \right)_{r=\Re^+} = 0. \quad (14)$$

Functions $\tilde{\eta}(r)$, being single-valued functions, are described by the Bessel functions of integer orders. An orthogonal spectrum of oscillations in a ferrite disk is obtained when one solves the characteristic equation for MS waves in an axially magnetized ferrite rod [arising from Eq. (14)]:

$$(-\mu)^{\frac{1}{2}} \frac{J'_\nu}{J_\nu} + \frac{K'_\nu}{K_\nu} = 0, \quad (15)$$

where $J_\nu, J'_\nu, K_\nu,$ and $K'_\nu$ are the values of the Bessel functions of an order $\nu$ and their derivatives (with respect to the argument) on a lateral cylindrical surface ($r = \Re$), together with the characteristic equation for MS waves in a normally magnetized ferrite slab:

$$\tan(\beta d) = -\frac{2\sqrt{-\mu}}{1+\mu}, \quad (16)$$

where $d$ is a disk thickness. Solutions for membrane MS-potential wave function $\tilde{\eta}(r,\theta)$, satisfying the second-order differential equation (12) with operator $\hat{G}_\perp$, we will conventionally call as *G-mode* magnetic-dipolar oscillations. The *G*-mode oscillations are characterized by the orthogonality relations with *energy eigenstates* [5]. The boundary condition (14) is a so-called essential boundary condition [6]. It is very important to note, however, that this boundary condition does not satisfy the condition of continuity of the magnetic flux density on a lateral surface of a disk [see Eq. (11)]. To settle the problem, one should impose a certain *boundary phase factor* which is described by singular edge wave functions. On a lateral border of a ferrite disk one has the following correspondence between a double-valued functions $\tilde{\varphi}$ and a single-valued functions $\tilde{\eta}$ [7]:

$$(\tilde{\varphi}_\pm)_{r=\Re^-} = \delta_\pm (\tilde{\eta})_{r=\Re^-}, \quad (17)$$

where

$$\delta_\pm \equiv f_\pm e^{-iq_\pm \theta} \quad (18)$$

is a double-valued function. The azimuth number $q_\pm$ is equal to $\pm \frac{1}{2}$ and for amplitudes we have $f_+ = -f_-$ and $|f_\pm| = 1$. Function $\delta_\pm$ changes its sign when $\theta$ is rotated by $2\pi$ so that $e^{-iq_\pm 2\pi} = -1$.

As a result, one has the energy-eigenstate spectrum of MDM oscillations with topological phases accumulated by the boundary wave function $\delta$. The topological effects become apparent through the integral fluxes of the pseudo-electric fields [7]. There should be the positive and negative fluxes corresponding to the counterclockwise and clockwise edge-function chiral rotations. The different-sign fluxes are inequivalent to avoid cancellation. Every MDM in a thin ferrite disk is characterized by a certain energy eigenstate and different-sign quantized fluxes of the pseudo-electric fields which are energetically degenerate. The spectral theory, developed based on orthogonal singlevalued membrane functions $\tilde{\eta}$ and topological magnetic currents, shows the magnetoelectric effect from a viewpoint of the Berry phase connection. From the theory of *G*-modes, it follows that a macroscopic-size ferrite disk may behave as a single quantum-like particle with the observable energy eigenstates and eigen electric moment properties of oscillating modes [5 – 7, 10 – 12]. It is very important to note that for the *G*-MDMs, the energy orthogonality relations one has only for spectra with respect to the DC bias magnetic field and at a constant oscillating frequency [6].

In the second approach, we formulate the spectral problem for MS-wave function with the boundary condition using continuity of the magnetic flux density on a lateral surface of a disk. There is a so-called natural boundary condition [6]. In this case, one may suppose that there will not be singular edge wave functions. The spectral problem is described by a differential-matrix-operator equation [5 – 7]

$$\left( \hat{L}_{\perp} - i\beta \, \hat{R} \right) \begin{pmatrix} \tilde{\vec{B}} \\ \tilde{\varphi} \end{pmatrix} = 0, \tag{19}$$

where $\tilde{\varphi}$ is a dimensionless membrane MS-potential wave function (different from membrane functions $\tilde{\eta}$), $\tilde{\vec{B}}$ is a dimensionless membrane function of a magnetic flux density. In Eq. (19), $\hat{L}_{\perp}$ is a differential-matrix operator:

$$\hat{L}_{\perp} \equiv \begin{pmatrix} (\ddot{\mu}_{\perp})^{-1} & \nabla_{\perp} \\ -\nabla_{\perp} \cdot & 0 \end{pmatrix}, \tag{20}$$

(subscript $\perp$ means correspondence with the in-plane, $r, \theta$, coordinates), $\beta$ is the MS-wave propagation constant along $z$ axis ($\varphi \equiv \tilde{\varphi} \, e^{-i\beta z}$, $\vec{B} = \tilde{\vec{B}} \, e^{-i\beta z}$), and $\hat{R}$ is a matrix:

$$\hat{R} \equiv \begin{pmatrix} 0 & \vec{e}_z \\ -\vec{e}_z & 0 \end{pmatrix}, \tag{21}$$

where $\vec{e}_z$ is a unit vector along $z$ axis. Continuity of functions $\tilde{\varphi}$ and $\tilde{\vec{B}}$ on a lateral surface of a ferrite disk give self-adjointess of operator $\hat{L}$. Solutions satisfying the first-order differential-matrix operator $\hat{L}$ we will conventionally call as *L-mode* magnetic-dipolar oscillations. There is an open question, however, about singlevaluedness of MS-potential wave functions for the *L*-mode solutions.

The spectral problem described by Eq. (19) presumes the possibility of separation of variables. In frames of such a presumption, the *L*-MDMs are normalized to the density of the power flow along *z* axis in an axially magnetized ferrite rod [5]. This mode normalization is not physically well justified, however. It is evident that for the MS-wave process in an axially magnetized ferrite rod described by functions $\tilde{\varphi}$ and $\tilde{\vec{B}}$, one has the longitudinal phase variations along $z$ axis (defined by the propagation constant $\beta$) together with azimuth phase variation [defined by the RHS term in Eq. (11)]. This results in appearance of helical waves with a combined effect of longitudinal and azimuth power flows. Proper integrable solutions for $L$-mode functions $\tilde{\varphi}$ should be obtained based on an analysis of the MDM propagation in a helical coordinate system. Such an analysis will give us evidence for the symmetry breaking effects of MDMs in a ferrite disk.

## 3.2. MDMs in a helical coordinate system

The helices are topologically nontrivial structures and the phase relationships for waves propagating in such structures could be very special. Unlike the Cartesian or cylindrical coordinate systems, in the helical system, two different types of solutions are admitted, one right-handed and one left-handed. Since the helical coordinates are nonorthogonal and curvilinear, different types of helical coordinate systems can be suggested. In our analysis we will use the Waldron coordinate system [37]. As an alternative helical coordinate system, we can point out, for example, the system proposed by Lin-Chung and Rajagopal [38]. Waldron showed [37] that the solution of the Helmholtz equation in a helical coordinate system can be reduced to the solution of the Bessel equation. With use of the Waldron coordinate system, Overfelt had got analytic exact solutions of the Laplace equation in a helical coordinate system with a reference to the helical Bessel functions and helical harmonics for static fields [39].

In the Waldron coordinate system, the pitch of the helix is fixed but the pitch angle is allowed to vary as a function of the radius. In cylindrical coordinates $(r,\theta,z)$, the reference surfaces, which are orthogonal, are given, respectively, by $r = const$, $\theta = const$, and $z = const$. In the Waldron helical system $(r,\phi,\zeta)$, we retain the family of cylinders $r = const$ with meaning unchanged, but instead of the parallel planes $z = const$, we use a family of helical surfaces given by $z = const + p\theta/2\pi$, where $p$ is the pitch. Fig. 1 shows the helical reference surfaces $z = p\theta/2\pi$ for the right-handed (a) and left-handed (b) helical coordinates. Coordinate $\zeta$ is measured parallel to coordinate $z$ from the reference surface $z = p\theta/2\pi$. The third coordinate surfaces is the set of planes $\theta = const$. We, however, use the azimuth coordinate $\phi$ instead of $\theta$. Coordinate $\phi$ is numerically equal to coordinate $\theta$, but whereas $\theta$ is measured in a plane $z = const$, $\phi$ is measured in a helical surface $\zeta = const$. Contrary to a cylindrical coordinate system, the helical coordinate system is not orthogonal.

Let us consider a wave process in a helical structure with a constant pitch $p$. Geometrically, a certain phase of the wave can reach a point $(r,\theta,z+p)$ from the point $(r,\theta,z)$ in two independent ways. In the first way, due to translation in the $\zeta$ direction at $r = const$ and $\theta = \phi = const$, and, in the second way, due to translation in $\phi$ at $r = const$ and $\zeta = const$. In other words, for any point $A$ with coordinates $(r,\phi,\zeta)$, the point $B$, being distant with a *period of a helix*, is characterized by coordinates $(r,\phi,\zeta+p)$ or by coordinates $(r,\phi+2\pi,\zeta)$. The regions between the surfaces $\zeta = np$ and $\zeta = (n+1)p$, for all integer numbers $n$, are continuous in a multiply connected space.

To analyze symmetry properties of magnetic-dipolar spin modes we consider now the MS-wave propagation in the Waldron right-handed and left-handed helical coordinate systems. For the first time, such studies we carried out in Refs. 40, 41. It was supposed [40, 41] that since in the Landau-Lifshitz equation there are opposite signs for vector products with respect to the right-handed and left-handed helical coordinate systems, the off-diagonal components of the permeability tensor should have different signs for the right-handed and left-handed helical MS waves. At the present stage of studies, such a statement in Refs. 40, 41 cast, however, certain doubts. It is clear that in the laboratory frame, the short-range local loop of the electron precession should be completely non-correlated with a character of the long-range helical MS-wave process in an entire ferrite sample. So one, certainly, can assume that in a helical coordinate system describing MS waves in a sample, the quantities of diagonal and off-diagonal components of the permeability tensor remain the same as in a cylindrical coordinate system and that signs of the off-diagonal components do not have any connection with the type of the helix. It means that the direction of the azimuth phase variation of the entire-sample helical wave and the direction of the local magnetization precession should be considered as separate notions. Since the direction of the magnetization precession is correlated with the direction of time, the above statement just shows that for a separate MS-potential helical wave the coordinate phase variation is independent on the time phase variation.

Let us suppose formally that there exist MS-wave helical solutions in a system characterized by a certain pitch $p$. For a DC magnetic field directed along $z$-axis, for both types of helices (right-handed or left-handed), one has the permeability tensor:

$$\ddot{\mu} = \begin{bmatrix} \mu & i\mu_a & 0 \\ -i\mu_a & \mu & 0 \\ 0 & 0 & 1 \end{bmatrix}, \tag{22}$$

where the components $\mu$ and $\mu_a$ can be found from Ref. 13. The signs of the tensor components are the same for both right-hand and left-hand helices, but with respect to the frequency and bias-field regions, the quantities of $\mu$ and $\mu_a$ may be positive or negative. For helical MS modes in an infinite axially magnetized ferrite rod, the components of the magnetic flux density are expressed by means of the components of the magnetic field in the Waldron helical coordinate system $(r,\phi,\zeta)$ [37, 40] as:

$$B_r = i\mu_a H_\phi (\cos\alpha_0)^{(R,L)} + \mu H_r,$$
$$B_\phi = \mu H_\phi - i\frac{\mu_a}{(\cos\alpha_0)^{(R,L)}} H_r, \tag{23}$$
$$B_\zeta = H_\zeta + (1-\mu)H_\phi (\sin\alpha_0)^{(R,L)} + i\mu_a H_r (\tan\alpha_0)^{(R,L)},$$

where superscripts $R$ and $L$ mean, respectively, right-handed and left-handed helical coordinate systems. The pitch angles are defined from the relations:

$$(\tan\alpha_0)^{(R)} \equiv \tan\alpha_0 \equiv \bar{p}/r \quad \text{and} \quad (\tan\alpha_0)^{(L)} = -\tan\alpha_0 = -\bar{p}/r, \tag{24}$$

where $\bar{p} = p/2\pi$. The quantities $\tan\alpha_0$ and $\bar{p}$ are assumed to be positive.

With representation of magnetic field as $\vec{H} = -\nabla \psi$ and with use of transformations in helical coordinates [37, 40], we rewrite Eqs. (23) as:

$$B_r = -\left[\mu \frac{\partial \psi}{\partial r} + i\mu_a \left(\frac{1}{r}\frac{\partial \psi}{\partial \phi} - (\tan \alpha_0)^{(R.L)} \frac{\partial \psi}{\partial \zeta}\right)\right],$$

$$B_\phi = -\frac{1}{(\cos \alpha_0)^{(R.L)}}\left[\mu\left(\frac{1}{r}\frac{\partial \psi}{\partial \phi} - (\tan \alpha_0)^{(R.L)} \frac{\partial \psi}{\partial \zeta}\right) - i\mu_a \frac{\partial \psi}{\partial r}\right], \quad (25)$$

$$B_\zeta = -(\tan \alpha_0)^{(R.L)}\left[\frac{2}{(\sin 2\alpha_0)^{(R.L)}}\frac{\partial \psi}{\partial \zeta} - \frac{1}{r}\frac{\partial \psi}{\partial \phi} + (1-\mu)\left(\frac{1}{r}\frac{\partial \psi}{\partial \phi} - (\tan \alpha_0)^{(R.L)} \frac{\partial \psi}{\partial \zeta}\right) + i\mu_a \frac{\partial \psi}{\partial r}\right].$$

Based on Waldron's equation for the divergence [37], we have

$$\nabla \cdot \vec{B} = \frac{1}{r}\frac{\partial}{\partial r}(rB_r) + \frac{\cos \alpha_0}{r}\frac{\partial B_\phi}{\partial \phi} + \frac{\partial B_\zeta}{\partial \zeta} = 0. \quad (26)$$

After some transformations we obtain the Walker equation [42] in helical coordinates:

$$\frac{\partial^2 \psi}{\partial r^2} + \frac{1}{r}\frac{\partial \psi}{\partial r} + \frac{1}{r^2}\frac{\partial^2 \psi}{\partial \phi^2} + \left(\frac{1}{\mu} + \tan^2 \alpha_0\right)\frac{\partial^2 \psi}{\partial \zeta^2} - 2\frac{1}{r}(\tan \alpha_0)^{(R.L)} \frac{\partial^2 \psi}{\partial \phi \partial \zeta} = 0. \quad (27)$$

Outside a ferrite region (where $\mu = 1$) Eq. (27) reduces to the Laplace equation in helical coordinates [37, 39]. Following Overfelt's approach [39], we assume that solutions of the Laplace and Walker equations are found as

$$\psi(r, \phi, \zeta) = R(r)P(\phi)Z(\zeta), \quad (28)$$

where

$$\begin{aligned} P(\phi) &\sim \exp(\pm iw\phi), \\ Z(\zeta) &\sim \exp(\pm i\beta\zeta). \end{aligned} \quad (29)$$

Here the quantities of wavenumbers $w$ and $\beta$ are assumed to be real and positive. For a chosen direction of $z$ axis, there are four solutions for the MS-potential wave function inside and outside a ferrite rod:

$$\begin{aligned} \psi^{(1)} &\sim e^{-iw\phi}e^{-i\beta\zeta}, \\ \psi^{(2)} &\sim e^{+iw\phi}e^{-i\beta\zeta}, \\ \psi^{(3)} &\sim e^{+iw\phi}e^{+i\beta\zeta}, \\ \psi^{(4)} &\sim e^{-iw\phi}e^{+i\beta\zeta}. \end{aligned} \quad (30)$$

We assume that the helical wave $\psi^{(1)}$ is the forward (propagating in a ferrite rod along the $z$ axis) right-hand-helix (FR) MS wave. For this wave we will take $(\tan\alpha_0)^{(R)} \equiv \tan\alpha_0 \equiv \bar{p}/r$. Other types of helical waves $(\psi^{(2)}, \psi^{(3)}, \psi^{(4)})$ will be considered with respect to the wave $\psi^{(1)}$. It is evident that the wave $\psi^{(2)}$ is the forward left-hand-helix (FL) wave, the wave $\psi^{(3)}$ is the backward (propagating in a ferrite rod oppositely to the direction of the $z$ axis) right-hand-helix (BR) wave, and the wave $\psi^{(4)}$ is the backward left-hand-helix (BL) wave. For the BR wave ($\psi^{(3)}$), there is $(\tan\alpha_0)^{(R)} = \tan\alpha_0 = \bar{p}/r$, and for the FL wave ($\psi^{(2)}$) and BL wave ($\psi^{(4)}$) there is $(\tan\alpha_0)^{(L)} = -\tan\alpha_0 = -\bar{p}/r$. As an example, Figs. 1 (a) and (b) illustrate, respectively, propagation of the FR helical wave $\psi^{(1)}$ and the FL helical wave $\psi^{(2)}$ in helical coordinate systems.

For helical waves in a ferrite rod of radius $\Re$, we have from Eq. (27):

$$\frac{\partial^2 \psi(r)}{\partial r^2} + \frac{1}{r}\frac{\partial \psi(r)}{\partial r} - \left[\frac{\beta^2}{\mu} + \frac{1}{r^2}(w - \bar{p}\beta)^2\right]\psi(r) = 0 \tag{31}$$

inside a ferrite rod $(r \leq \Re)$ and

$$\frac{\partial^2 \psi(r)}{\partial r^2} + \frac{1}{r}\frac{\partial \psi(r)}{\partial r} - \left[\beta^2 + \frac{1}{r^2}(w - \bar{p}\beta)^2\right]\psi(r) = 0 \tag{32}$$

outside a ferrite rod $(r \geq \Re)$. Physically acceptable solutions for Eqs. (31) and (32) are possible only for negative quantities $\mu$. Inside a ferrite region ($r \leq \Re$) the solutions are in the form:

$$\begin{aligned}
\psi^{(1)} &= a_1 J_{(w-\bar{p}\beta)}\left[(-\mu)^{1/2}\beta r\right] e^{-iw\phi} e^{-i\beta\zeta}, \\
\psi^{(2)} &= a_2 J_{(w-\bar{p}\beta)}\left[(-\mu)^{1/2}\beta r\right] e^{+iw\phi} e^{-i\beta\zeta}, \\
\psi^{(3)} &= a_3 J_{(w-\bar{p}\beta)}\left[(-\mu)^{1/2}\beta r\right] e^{+iw\phi} e^{+i\beta\zeta}, \\
\psi^{(4)} &= a_4 J_{(w-\bar{p}\beta)}\left[(-\mu)^{1/2}\beta r\right] e^{-iw\phi} e^{+i\beta\zeta}.
\end{aligned} \tag{33}$$

For an outside region ($r \geq \Re$) one has:

$$\begin{aligned}
\psi^{(1)} &= b_1 K_{(w-\bar{p}\beta)}(\beta r) e^{-iw\phi} e^{-i\beta\zeta}, \\
\psi^{(2)} &= b_2 K_{(w-\bar{p}\beta)}(\beta r) e^{+iw\phi} e^{-i\beta\zeta}, \\
\psi^{(3)} &= b_3 K_{(w-\bar{p}\beta)}(\beta r) e^{+iw\phi} e^{+i\beta\zeta}, \\
\psi^{(4)} &= b_4 K_{(w-\bar{p}\beta)}(\beta r) e^{-iw\phi} e^{+i\beta\zeta}.
\end{aligned} \tag{34}$$

Here $J$ and $K$ are Bessel functions of real and imaginary arguments, respectively. Coefficients $a_{1,2,3,4}$ and $b_{1,2,3,4}$ are amplitude coefficients.

Now we can obtain proper equations for magnetic flux density components of helical waves. For the FR and BR waves there are

$$B_r^{(1,3)} = -\left[\mu\frac{\partial\psi}{\partial r} + i\mu_a\left(\frac{1}{r}\frac{\partial\psi}{\partial\phi} - \tan\alpha_0\frac{\partial\psi}{\partial\zeta}\right)\right],$$

$$B_\phi^{(1,3)} = -\frac{1}{\cos\alpha_0}\left[\mu\left(\frac{1}{r}\frac{\partial\psi}{\partial\phi} - \tan\alpha_0\frac{\partial\psi}{\partial\zeta}\right) - i\mu_a\frac{\partial\psi}{\partial r}\right], \quad (35)$$

$$B_\zeta^{(1,3)} = -\tan\alpha_0\left[\frac{2}{\sin 2\alpha_0}\frac{\partial\psi}{\partial\zeta} - \frac{1}{r}\frac{\partial\psi}{\partial\phi} + (1-\mu)\left(\frac{1}{r}\frac{\partial\psi}{\partial\phi} - \tan\alpha_0\frac{\partial\psi}{\partial\zeta}\right) + i\mu_a\frac{\partial\psi}{\partial r}\right].$$

and for the FL and BL waves we have

$$B_r^{(2,4)} = -\left[\mu\frac{\partial\psi}{\partial r} + i\mu_a\left(\frac{1}{r}\frac{\partial\psi}{\partial\phi} + \tan\alpha_0\frac{\partial\psi}{\partial\zeta}\right)\right],$$

$$B_\phi^{(2,4)} = -\frac{1}{\cos\alpha_0}\left[\mu\left(\frac{1}{r}\frac{\partial\psi}{\partial\phi} + \tan\alpha_0\frac{\partial\psi}{\partial\zeta}\right) - i\mu_a\frac{\partial\psi}{\partial r}\right], \quad (36)$$

$$B_\zeta^{(2,4)} = \tan\alpha_0\left[-\frac{2}{\sin 2\alpha_0}\frac{\partial\psi}{\partial\zeta} - \frac{1}{r}\frac{\partial\psi}{\partial\phi} + (1-\mu)\left(\frac{1}{r}\frac{\partial\psi}{\partial\phi} + \tan\alpha_0\frac{\partial\psi}{\partial\zeta}\right) + i\mu_a\frac{\partial\psi}{\partial r}\right].$$

The above four helical waves should be considered as components of the MS-potential function and components of the magnetic flux density. So we have the following four-component functions:

$$[\psi] = \begin{pmatrix}\psi^{(1)}\\\psi^{(2)}\\\psi^{(3)}\\\psi^{(4)}\end{pmatrix}, \quad [B_r] = \begin{pmatrix}B_r^{(1)}\\B_r^{(2)}\\B_r^{(3)}\\B_r^{(4)}\end{pmatrix}, \quad [B_\phi] = \begin{pmatrix}B_\phi^{(1)}\\B_\phi^{(2)}\\B_\phi^{(3)}\\B_\phi^{(4)}\end{pmatrix}, \quad [B_\zeta] = \begin{pmatrix}B_\zeta^{(1)}\\B_\zeta^{(2)}\\B_\zeta^{(3)}\\B_\zeta^{(4)}\end{pmatrix}. \quad (37)$$

Since $B_z = B_\zeta + B_\phi\sin\alpha_0$ [37], one has from Eqs. (35) and (36) after some algebraic transformations:

$$B_z^{(1,2,3,4)} = -\frac{\partial\psi^{(1,2,3,4)}}{\partial\zeta} = -\frac{\partial\psi^{(1,2,3,4)}}{\partial z}. \quad (38)$$

One can rewrite this equation as:

$$[B_z] = -\begin{pmatrix} \nabla_\parallel \psi^{(1)} \\ \nabla_\parallel \psi^{(2)} \\ \nabla_\parallel \psi^{(3)} \\ \nabla_\parallel \psi^{(4)} \end{pmatrix}. \tag{39}$$

On a cylindrical surface of a ferrite rod of radius $\mathfrak{R}$ we have the boundary conditions:

$$(\psi)_{r=\mathfrak{R}^-} = (\psi)_{r=\mathfrak{R}^+} \quad \text{and} \quad (B_r)_{r=\mathfrak{R}^-} = (B_r)_{r=\mathfrak{R}^+}. \tag{40}$$

In a general form, the boundary condition for radial components of the magnetic flux density [see Eqs. (35) and (36)] can be written as

$$\left[\mu\left(\frac{\partial \psi}{\partial r}\right) + i\mu_a \frac{1}{\mathfrak{R}}\left(\frac{\partial \psi}{\partial \phi} \mp \bar{p}\frac{\partial \psi}{\partial \zeta}\right)\right]_{r=\mathfrak{R}^-} = \left(\frac{\partial \psi}{\partial r}\right)_{r=\mathfrak{R}^+}. \tag{41}$$

Based on the above Bessel equations and boundary conditions one obtains, as a result, characteristic equations for helical MS waves in a ferrite rod. For helical modes $\psi^{(1)}$ and $\psi^{(4)}$ there is a characteristic equation in a form:

$$(-\mu)^{1/2} \frac{J'_{(w-\bar{p}\beta)}}{J_{(w-\bar{p}\beta)}} + \frac{K'_{(w-\bar{p}\beta)}}{K_{(w-\bar{p}\beta)}} - \frac{\mu_a(w-\bar{p}\beta)}{\beta \mathfrak{R}} = 0. \tag{42}$$

For helical modes $\psi^{(2)}$ and $\psi^{(3)}$ one has:

$$(-\mu)^{1/2} \frac{J'_{(w-\bar{p}\beta)}}{J_{(w-\bar{p}\beta)}} + \frac{K'_{(w-\bar{p}\beta)}}{K_{(w-\bar{p}\beta)}} + \frac{\mu_a(w-\bar{p}\beta)}{\beta \mathfrak{R}} = 0. \tag{43}$$

In these equations, the prime denotes differentiation with respect to the argument.

As we discussed above, for any point *A* with coordinates $(r,\phi,\zeta)$, the point *B*, being distant with a period of a helix, is characterized by coordinates $(r,\phi,\zeta+p)$ or by coordinates $(r,\phi+2\pi,\zeta)$. It means that for any *separate* helical wave there is a trivial relation: $w = \bar{p}\beta$. For this relation, Eqs. (42) and (43) are reduced to the equation

$$(-\mu)^{1/2} \frac{J'_0}{J_0} + \frac{K'_0}{K_0} = 0. \tag{44}$$

This equation has singlevalued solutions, but does not have any sense for our studies. It is evident that in a cylindrical coordinate system, Eq. (44) does not reflect the azimuth phase variation appearing due to the boundary conditions for the magnetic flux density.

In an analysis of the MDM oscillation resonances any separate helix is of no interest to us, while combination of two helices may have a physical meaning. From the characteristic equation (42), it becomes clear that we have the same quantities of $w - \bar{p}\beta$ for helical modes $\psi^{(1)}$ and $\psi^{(4)}$. It means that phases of these two helical waves (propagating in opposite directions of *z*-axis, but at the same direction of azimuth rotation) are reciprocally correlated and may be linked together. Two constituting objects, $\psi^{(1)}$ and $\psi^{(4)}$, with non-local connection are considered as entangled helical waves. A similar phase correlation (entanglement) for helical waves $\psi^{(2)}$ and $\psi^{(3)}$ one has from Eq. (43). At the same time, waves $\psi^{(1)}$ and $\psi^{(3)}$ as well as waves $\psi^{(2)}$ and $\psi^{(4)}$ are not linked together (not entangled). One can see that at a given radius *r* for certain phase correlations between helical waves $\psi^{(1)}$ and $\psi^{(4)}$ (as well as for helical waves $\psi^{(2)}$ and $\psi^{(3)}$), there is the possibility to obtain interference along *z*-axis. So for certain phase correlations between helical-mode MS-potential wave functions, there could be nodes and bulges in some points along *z*-axis. It is necessary to note, however, that it is not a "classical" example of a standing wave. The possible nodes and bulges for the MS-potential wave function along *z*-axis are due to the entanglement.

For reciprocally correlated (entangled) waves, having the same quantities of $w - \bar{p}\beta$, one can consider the *closed-loop* phase run of the *double-helix* resonance. For waves $\psi^{(1)}$ and $\psi^{(4)}$, the closed-loop phase way has two parts: the forward part of the way is along the FR wave $\psi^{(1)}$ and the backward part of the way – along the BL wave $\psi^{(4)}$. To have such a closed-loop phase way, we may assume that

$$w^{(1)} = w^{(4)}, \quad \beta^{(1)} = \beta^{(4)}, \quad \text{and} \quad \bar{p}^{(1)} = \bar{p}^{(4)}. \tag{45}$$

Similarly, we may have the closed-loop phase way for waves $\psi^{(2)}$ and $\psi^{(3)}$ if we assume that

$$w^{(2)} = w^{(3)}, \quad \beta^{(2)} = \beta^{(3)}, \quad \text{and} \quad \bar{p}^{(2)} = \bar{p}^{(3)}. \tag{46}$$

Eqs. (45) and (46) describe, in fact, the resonance conditions. One resonance state is due to the $\psi^{(1)} \leftrightarrow \psi^{(4)}$ phase correlation and another resonance state is due to the $\psi^{(2)} \leftrightarrow \psi^{(3)}$ phase correlation. The resonance $\psi^{(1)} \leftrightarrow \psi^{(4)}$, being characterized by the right-hand rotation (with respect to a bias magnetic field directed along *z*-axis) of a composition of helices, we will conventionally call as the (+) resonance, while the resonance $\psi^{(2)} \leftrightarrow \psi^{(3)}$ with the left-hand rotation of a helix composition we will conventionally call as the (–) resonance. In a case of the (+) double-helix resonance, the azimuth phase over-running of the MS-potential wave functions is in a correspondence with the right-hand resonance rotation of magnetization in a ferrite magnetized along *z*-axis by a DC magnetic field [13].

The solutions for the (+) and (–) resonances can be represented, respectively, as

$$[\psi^{(+)}] = \begin{pmatrix} \psi^{(1)} \\ \psi^{(4)} \end{pmatrix} \quad \text{and} \quad [\psi^{(-)}] = \begin{pmatrix} \psi^{(2)} \\ \psi^{(3)} \end{pmatrix} \tag{47}$$

It is also necessary to note that from Eqs. (35) and (36) one has

$$B_r^{(1)} = B_r^{(4)} \quad \text{and} \quad B_r^{(2)} = B_r^{(3)}. \tag{48}$$

At the same time,

$$B_r^{(1)} \neq B_r^{(2)}, \; B_\theta^{(1)} \neq B_\theta^{(2)} \quad \text{and} \quad B_r^{(3)} \neq B_r^{(4)}, \; B_\theta^{(3)} \neq B_\theta^{(4)} \; . \tag{49}$$

It is evident that for a smooth infinite ferrite rod, no concrete pitch parameter exists and so all the above analysis of the helical mode propagation bears a formal character and does not have a physical meaning. The problem, however, acquires a real physical meaning in a case of a restricted ferrite-rod waveguide section – a ferrite disk. In a disk, the quantity of pitch $\bar{p}$ is determined by the virtual "reflection" planes – the planes perpendicular to z-axis, where the nodes or bulges of helical MS-potential wave functions can be located. We introduce now an effective disk thickness $d_{eff}$, so that the planes $z = -d^{eff}/2$ and $z = d^{eff}/2$ are virtual "reflection" planes. It is necessary to note that the helical wave $\psi^{(4)}$ is not a really reflected wave with respect to the wave $\psi^{(1)}$ and, vice versa, the wave $\psi^{(1)}$ is not a result of reflection of the incident wave $\psi^{(4)}$. The same assertion one has to express with respect to helical waves $\psi^{(2)}$ and $\psi^{(3)}$. The resonant states are the states of the phase correlated (entangled) helices. Since for the entangled states $\psi^{(1)} \leftrightarrow \psi^{(4)}$ and $\psi^{(2)} \leftrightarrow \psi^{(3)}$ there are different pitches [see Eqs. (45) and (46)], we have to consider different quantities $d_{eff}^{(1\leftrightarrow 4)}$ and $d_{eff}^{(2\leftrightarrow 3)}$ for the (+) and (–) resonances, respectively. For the entangled (double-helix) states, $\psi^{(1)} \leftrightarrow \psi^{(4)}$ or $\psi^{(2)} \leftrightarrow \psi^{(3)}$, there should be phase over-running for the azimuth coordinate, but no phase over-running for the MS-potential wave function along z-axis.

Let us consider the case of for the (+) double-helix resonance, $\psi^{(1)} \leftrightarrow \psi^{(4)}$, when the closed loop appears in a ferrite-rod section with $d_{eff}^{(1\leftrightarrow 4)}$ equal to $p^{(+)}/2$. This resonance state is illustrated in Fig. 2. For every separate helical wave, the phase shift between points a and b (see. Fig. 2) is equal to $\pi$. We suppose now that along z-axis a mutual phase shift between the waves $\psi^{(1)}$ and $\psi^{(4)}$ is also $\pi$ so that the MS potential $\psi = \psi^{(1)} + \psi^{(4)} = 0$ in points a and b. The planes $z = -d^{eff}/2$ and $z = d^{eff}/2$ can be characterized as the "magnetic walls". At the same time, the waves $\psi^{(1)}$ and $\psi^{(4)}$ are in phase with respect to the azimuth coordinate. One has, as a result, the condition that all the composition of waves $\psi^{(1)}$ and $\psi^{(4)}$ is running azimuthally counterclockwise without any phase over-running along z-axis. Since in the Waldron helical coordinate system we retain the family of cylinders $r = const$ with meaning unchanged with respect to the cylindrical coordinate system, radial variations in a disk described by a composition of two helical waves can be expressed in cylindrical coordinates with replacing the Bessel function order $w^{(1,4)} - \bar{p}^{(1,4)}\beta^{(1,4)}$ in Eq. (42) by a certain quantity $\nu^{(+)}$. In such a case, we can introduce a notion of an equivalent membrane function (EMF). For the case shown in Fig. 2, the azimuth phase over-running for the EMF is characterized by the azimuth wave number $\nu^{(+)} = 1$. It is worth noting that $\nu^{(+)}$ shows the geometrical phase variation, while $w^{(1,4)}$ characterizes the dynamical phase variation. When a composition of waves $\psi^{(1)} \leftrightarrow \psi^{(4)}$ acquires a geometrical phase over-running of $2\pi$ during a period, every separate wave, $\psi^{(1)}$ or $\psi^{(4)}$, has a dynamical phase over-running of $\pi$ and so behaves as a double-valued function. There can be different combinations of the $\psi^{(1)}$ and $\psi^{(4)}$ double-valued-function branches. Fig. 3 illustrates the (+) double-helix resonance, $\psi^{(1)} \leftrightarrow \psi^{(4)}$, when $d_{eff}^{(1\leftrightarrow 4)}$ is equal to $p^{(+)}$. The MS potential $\psi = \psi^{(1)} + \psi^{(4)} = 0$ in points *a*, *b*, and *c*. In this case, as well, we have the geometrical phase variation with the EMF characterized by the azimuth wave number $\nu^{(+)} = 1$. One can easily picture the situation of $\nu^{(+)} = 2$. In this case, in a disk plane one will see two double-helix loops geometrically shifted to $90°$. Based on the above consideration for the (+) double-helix resonance, we can rewrite Eq. (42) in a form:

$$(-\mu)^{1/2}\left(\frac{J'_{\nu^{(+)}}}{J_{\nu^{(+)}}}\right)_{r=\Re} + \left(\frac{K'_{\nu^{(+)}}}{K_{\nu^{(+)}}}\right)_{r=\Re} - \frac{\nu^{(+)}\mu_a}{\beta^{(+)}\Re} = 0, \qquad (50)$$

where $\nu^{(+)} = 1, 2, 3,...$ and $\beta^{(+)} \equiv \beta^{(1)} = \beta^{(4)}$. This is an equation for the EMF radius variations.

A similar analysis one can make for the (–) double-helix resonance, $\psi^{(2)} \leftrightarrow \psi^{(3)}$. Fig. 4 shows such a resonant state for $d_{eff}^{(2\leftrightarrow 3)} = p^{(-)}/2$. Different combinations of the $\psi^{(2)}$ and $\psi^{(3)}$ double-valued-function branches are possible in a case of the (–) resonance like, for example, to those discussed above for the (+) resonance. In the (–) double-helix resonance one can also introduce a notion of the EMF and so rewrite Eq. (43) as

$$(-\mu)^{1/2}\left(\frac{J'_{\nu^{(-)}}}{J_{\nu^{(-)}}}\right)_{r=\Re} + \left(\frac{K'_{\nu^{(-)}}}{K_{\nu^{(-)}}}\right)_{r=\Re} + \frac{\nu^{(-)}\mu_a}{\beta^{(-)}\Re} = 0, \qquad (51)$$

where $\nu^{(-)} = 1, 2, 3,...$ is the azimuth wave number for EMFs at the (–) resonance and $\beta^{(-)} \equiv \beta^{(2)} = \beta^{(3)}$.

It becomes evident, however, that an introduction of the notion of the EMF (and so the possibility to reduce a problem to description in cylindrical coordinates) has a physical meaning only for the (+) double-helix resonance, but not for the (–) double-helix resonance. When one substitutes the solution in the form

$$\tilde{\varphi}^{(+)} \propto J_{\nu^{(+)}}\left(\frac{\beta^{(+)} r}{\sqrt{-\mu}}\right) e^{-j\nu^{(+)}\theta} \qquad (52)$$

into Eq. (11), one obtains Eq. (50). It means that solution in a form of Eq. (52) is described by the integer-order Bessel function in cylindrical coordinates. This is the singlevalued solution which reflects the azimuth phase variation appearing because of the boundary conditions for the magnetic flux density. As a result, we have possibility for separation of variables. The problem becomes integrable. Regarding Eq. (51), it is evident that there are no proper EMF solutions correlated with the boundary conditions for the magnetic flux density.

A real ferrite disk is an open thin-film structure with a small thickness/diameter ratio. For a real thin-film ferrite disk one has a small thickness *d*, so that $\beta^{(+)}d \ll \beta^{(+)}d_{eff}^{(1\leftrightarrow 4)}$. It means that the virtual "reflection" planes for helical modes can be found in free space regions far above and below a real disk. These "reflection" planes are, in fact, the mapping planes. Fig. 5 illustrates the (+) resonance in a real thin-film disk for the case when $d_{eff}^{(1\leftrightarrow 4)} = p^{(+)}/2$ and $v^{(+)} = 1$. Fig. 6 shows the (+) resonance in a real thin-film disk for the case when $d_{eff}^{(1\leftrightarrow 4)} = p^{(+)}$ and $v^{(+)} = 1$. For an open thin-film structure, in a case of the (+) resonance, one can use the method of separation of variables as well. This becomes evident from the following arguments: (a) the possibility to introduce the EMFs inside a ferrite disk and (b) the fact that the boundary conditions on the planes $z = \pm d/2$, demanding continuity of $\psi$ and $B_z$, are the same for every type of a helical wave. For helical waves $\psi^{(1)}$ and $\psi^{(4)}$, the boundary conditions on plane surfaces of a ferrite disk are the following:

$$\psi^{(1)}\Big|_{z=\left|\frac{d}{2}\right|^+} = \psi^{(1)}\Big|_{z=\left|\frac{d}{2}\right|^-}, \qquad \left(B_z\right)^{(1)}\Big|_{z=\left|\frac{d}{2}\right|^+} = \left(B_z\right)^{(1)}\Big|_{z=\left|\frac{d}{2}\right|^-}, \qquad (53a)$$

$$\psi^{(4)}\Big|_{z=\left|\frac{d}{2}\right|^+} = \psi^{(4)}\Big|_{z=\left|\frac{d}{2}\right|^-}, \qquad \left(B_z\right)^{(4)}\Big|_{z=\left|\frac{d}{2}\right|^+} = \left(B_z\right)^{(4)}\Big|_{z=\left|\frac{d}{2}\right|^-}. \qquad (53b)$$

From Eq. (38) it evidently follows that for both helical waves, $\psi^{(1)}$ and $\psi^{(4)}$, there are the same conditions for continuity of $B_z$.

In an assumption of separation of variables, there are exponentially descending solutions along $z$ axis for MS-potential functions in regions above and below a disk. For $z \leq -d/2$, $z \geq d/2$, and $r \leq \Re$ we describe the MS-potential function by the Bessel equation:

$$\frac{\partial^2 \psi^{(+)}(r)}{\partial r^2} + \frac{1}{r}\frac{\partial \psi^{(+)}(r)}{\partial r} + \left[\left(\alpha^{(+)}\right)^2 - \frac{\left(v^{(+)}\right)^2}{r^2}\right]\psi^{(+)}(r) = 0, \qquad (54)$$

where $\alpha^{(+)}$ is a real quantity. The solutions are

$$\psi^{(+)} = (f_1)^{(+)} e^{-iv^{(+)}\theta} e^{-\alpha^{(+)}\left(z - \frac{1}{2}d\right)} \qquad (55)$$

for $z \geq d/2$ and

$$\psi^{(+)} = (f_2)^{(+)} e^{-iv^{(+)}\theta} e^{\alpha^{(+)}\left(z + \frac{1}{2}d\right)}. \qquad (56)$$

for $z \leq -d/2$. Coefficients $f_1$ and $f_2$ are the amplitude coefficients. Similarly to the method used in Refs. [5, 6] one obtains, as a result, a system of two equations. There are the ferrite-rod equation (50) and the ferrite-slab equation (see Refs. [5, 6]):

$$\tan\left(\beta^{(+)}d\right) = -\frac{2\sqrt{-\mu}}{1+\mu}. \qquad (57)$$

Eq. (57) for *L*-modes is similar to Eq. (16) used for *G*-modes.

## 4. The vortex structures and azimuthally running waves of magnetization for *L*-MDMs

The above analysis of the MDM propagation in a helical coordinate system gives a proper justification for the *L*-mode membrane functions $\tilde{\varphi}$ and gives possibility to obtain integrable solutions for *L*-modes in a cylindrical coordinate system. While *G*-modes are orthogonal ones and for these modes one has the quantum-like solutions with energy eigenstates [5 – 7], the *L*-modes are non-orthogonal modes and solutions for these modes are classical-like. For *L*-modes one can observe the field structures in any local point of a ferrite disk [8, 9].

Because of separation of variables in a cylindrical coordinate system, the MS-potential wave function for *L*-mode can be written as

$$\psi_{\nu^{(+)},q} = C \xi_{\nu^{(+)},q}(z) \tilde{\varphi}_{\nu^{(+)},q}(r,\theta), \tag{58}$$

where $\tilde{\varphi}_{\nu^{(+)},q}(r,\theta)$ is a dimensionless effective membrane function defined as a solution of Eq. (50), $\xi_{\nu^{(+)},q}(z)$ is an amplitude factor defined as a solution of Eq. (57), and *C* is a dimensional coefficient, $\nu^{(+)} = 1, 2, 3,...$ is the Bessel-function order, and $q = 1, 2, 3,...$ is the number of zeros of the Bessel functions corresponding to different radial variations of the EMF. In cylindrical coordinates, the azimuth and *z*-axis variations of the wave function are not mutually correlated and one can impose the boundary conditions independently for the longitudinal *z* and the in-plane $r,\theta$ coordinates. In our studies we will consider the case of $d_{eff}^{(1\leftrightarrow 4)} = p^{(+)}/2$ and $\nu^{(+)} = 1$. Inside a ferrite disk ($r \leq \Re$, $-d/2 \leq z \leq d/2$) one represents the MS-potential function as

$$\psi(r,\theta,z,t) = C J_1\left(\frac{\beta^{(+)} r}{\sqrt{-\mu}}\right)\left(\cos\beta^{(+)} z + \frac{1}{\sqrt{-\mu}}\sin\beta^{(+)} z\right)e^{-i\theta}e^{i\omega t}. \tag{59}$$

Outside a ferrite disk, for $r \geq \Re$ and $-d/2 \leq z \leq d/2$, one has

$$\psi(r,\theta,z,t) = C K_1(\beta^{(+)} r)\left(\cos\beta^{(+)} z + \frac{1}{\sqrt{-\mu}}\sin\beta^{(+)} z\right)e^{-i\theta}e^{i\omega t}. \tag{60}$$

These solutions show the *azimuthally-propagating-wave* behavior for EMFs. There are *rotationally non-symmetric* waves.

The normalized "thickness" functions $\xi(z)$ for the 1st (*q* = 1) and 2nd (*q* = 2) MDMs are shown in Fig. 7. Fig. 8 illustrates the effective membrane functions $\tilde{\varphi}$ for the 1st (*q* = 1) MDM at different time phases. The calculations were made for a ferrite disk with the following material parameters: the saturation magnetization is $4\pi M_s = 1880$ G and the linewidth is $\Delta H = 0.8$ Oe. The disk diameter is $D = 3$ mm and the thickness $t = 0.05$ mm. The disk is normally magnetized by the bias magnetic field $H_0 = 4900$ Oe. Generally, these data correspond to the sample parameters used in microwave experiments [10 – 12]. The MDM resonance frequencies, obtained from solutions of Eqs. (50) and (57), are: *f* = 8.548 GHz for the 1st MDM and *f* = 8.667 GHz for the 2nd MDM.

Based the known MS-potential function, one defines the magnetic field $\vec{H} = -\vec{\nabla}\psi$ for every $L$-mode. One can obtain also components of the magnetization $\vec{m}$ as

$$\vec{m} = -\vec{\tilde{\chi}} \cdot \nabla \psi, \qquad (61)$$

where

$$\vec{\tilde{\chi}} = \begin{bmatrix} \chi & i\chi_a & 0 \\ -i\chi_a & \chi & 0 \\ 0 & 0 & 0 \end{bmatrix} \qquad (62)$$

is the magnetic susceptibility tensor, and components of the magnetic flux density as

$$\vec{B} = -\vec{\tilde{\mu}} \cdot \nabla \psi, \qquad (67)$$

where the permeability tensor is defined by Eq. (22). Parameters of tensors $\vec{\tilde{\chi}}$ and $\vec{\tilde{\mu}}$ are found from Ref. [13].

Helical wavefronts of MS-potential wavefunctions presume an azimuth component of the power flow density. This azimuth component can be found from an analysis of the power flow density for $L$-modes. In a general representation, for monochromatic MS-wave processes with time variation $\sim e^{i\omega t}$, the power flow density for a certain magnetic-dipolar mode $n$ (in Gaussian units) is expressed as [8]:

$$\vec{p}_q = \frac{i\omega}{16\pi}\left(\psi_q^* \vec{B}_q - \psi_q \vec{B}_q^*\right). \qquad (68)$$

It was shown [8] that there exist only an azimuth component of the power flow density for EMF. For $\nu^{(+)} = 1$, taking into account that $\tilde{\varphi}_q = \tilde{\varphi}_q(r)\tilde{\varphi}_n(\theta)$, where $\tilde{\varphi}_q(\theta) \sim e^{-i\theta}$, one obtains

$$\left(p_q(r,z)\right)_\theta = \frac{\tilde{\varphi}_q(r)}{8\pi} \omega C^2 \left(\xi_q(z)\right)^2 \left[-\tilde{\varphi}_q(r)\frac{\mu}{r} - \mu_a \frac{\partial \tilde{\varphi}_q(r)}{\partial r}\right]. \qquad (69)$$

This is a non-zero circulation quantity around a circle $2\pi r$. An amplitude of a MS-potential function is equal to zero at $r = 0$. For a scalar wave function, this presumes the Nye and Berry phase singularity [43]. Circulating quantities $\left(p_q(r,z)\right)_\theta$ are the MDM power-flow-density vortices with cores at the disk center. At a vortex center amplitude of $(p_q)_\theta$ is equal to zero. As an example, in Fig. 9 one can see the picture of the power flow density distribution for the 1$^{st}$ ($q$ = 1) $L$-MDM calculated based on Eq. (69).

In a ferrite sample with MDM oscillations, one has non-homogeneous precession of magnetization. The azimuthally-propagating-wave behavior for the MS-potential EMFs in a quasi-2D ferrite disk necessarily presumes the azimuth waves for magnetization $\vec{m}$. Figs. 10 and 11 show galleries of magnetization $\vec{m}$ at different time phases for the 1$^{st}$ ($q$ = 1) and 2$^{nd}$ ($q$ = 2) MDMs, respectively. It is evident that for a given radius $r$ there is the phase-running variation of magnetization with respect to azimuth angle $\theta$. At the same time, for a given azimuth angle $\theta$, the magnetization vectors are in phase or $180°$ out of phase in the radial direction. Figs. 12 and 13 illustrates this statement more explicitly for the 1st ($q = 1$) and 2nd ($q = 2$) MDMs, respectively. Cyclic evolution of magnetization gives the geometric phase. For MDMs in a ferrite disk, one can see the precession dynamics in correlation with the cyclic geometrical phase evolution of magnetization $\vec{m}$. At a given time phase $\omega_{res}^{(q)} t$, where $\omega_{res}^{(q)}$ is the resonance frequency of the $q$-th MDM, the precessing magnetization vector have different phases (with respect to the unit azimuth vector $\vec{e}_\theta$) in different parts of a sample. When (for a given radius inside a disk and a given time phase $\omega_{res}^{(q)} t$) an azimuth angle $\theta$ varies from 0 to $2\pi$, the magnetization vector accomplishes the $2\pi$ geometric-phase rotation. Because of magnetic ordering in a ferrite, the collective states of dipolar interacting precessing spins are characterized by strong spin correlations. Due to dipolar interactions, the precessing spins may exhibit macroscopic quantum coherence behavior allowing the existence of long-range ordered phases. This phase ordering assumes that different spin states in the whole-structure MDM share the same spatial long-range MS-potential wave function. The macroscopic magnetization $\vec{m}$ in a given point inside a quasi-2D ferrite disk is a macroscopically distinct state and because of a superposition of such macroscopically distinct states, every MDM can be considered as a macroscopically entangled quantum state. Solid-state spin systems have been proposed as possible candidates for large scale realizations of quantum entanglement [14]. There is increasing interest in combined phenomena of the geometric phase and the entanglement of a system [15, 16, 36]. The geometrical phase evolution of magnetization $\vec{m}$ in a quasi-2D ferrite disk is the Berry phase implemented by a rotating magnetic field of a MDM. Such a rotating magnetic field is an attribute of MDM oscillations in a disk [8, 9].

The observed azimuthally running waves of magnetization are rotating states with lack of inversion symmetry dependent on the spin rotational symmetry. The antisymmetric dipole-dipole interaction between two magnetic moments results in the magnon modes which display chirality. Both contributions – the precession dynamics and the geometrical phase evolution – of the magnetization field cannot be measured separately. This differs from the known magnetic system where in the crystal lattice with lack of inversion symmetry the chiral effects are independent of the spin rotational symmetry [44]. The hallmark of chirality is that it is exhibited by systems existing in two distinct states that are time-invariant and interconverted by space inversion. In our case, the chiral states are not time-invariant. Evidence for chirality is given by the double-helix behavior of MDM oscillations in a thin-film ferrite disk. For time reversal (which means an opposite direction of a bias magnetic field) the double-helix resonance will take place with another set of helical waves. Concerning the above statement on chirality of the MDMs, one can argue that in a usual Faraday effect in an unbounded ferrite medium there are also two rotating waves with the forward-back propagation which can be considered as waves with lack of inversion symmetry. The main aspect, however, is that these waves are not helical modes since no concrete pitch is determined.

It is evident that in a disk center, the magnetization of a MDM has a maximum. So one does not have magnetization vortices for MDM oscillations (see e.g. Ref. [4]). The azimuthally running waves of magnetization are, in fact, circulating magnetization currents. It becomes evident that such circulating magnetization currents with the inversion symmetry violation should be accompanied by electric fluxes piercing the magnetic current rings. Such electric fluxes studied recently in Ref. [7], give a new insight into physics of interactions between MDM ferrite disks [45].

## 5. The spectra and symmetry breaking effects for MDM oscillations

As we noted above, for a bias magnetic field directed along $z$ axis, only for the (+) double-helix resonance, $\psi^{(1)} \leftrightarrow \psi^{(4)}$, one can introduce the EMFs and consider spectral properties of $L$-MDMs in a cylindrical coordinate system. No solutions in a cylindrical coordinate system can be obtained for the (–) double-helix resonance, $\psi^{(2)} \leftrightarrow \psi^{(3)}$. Nevertheless, such a resonance exists. A proper solution for this resonance can be obtained only in a helical coordinate system.

When a bias magnetic field is directed contrarily to $z$ axis, Eq. (50) describes the (+) double-helix resonance, which is now due to the $\psi^{(2)} \leftrightarrow \psi^{(3)}$ phase correlated (entangled) helices. In this case $\nu^{(+)} = 1, 2, 3,...$ as well, but $\beta^{(+)} \equiv \beta^{(2)} = \beta^{(3)}$. Eq. (50) is now an equation for the EMF radius variations in a cylindrical coordinate system for the $\psi^{(2)} \leftrightarrow \psi^{(3)}$ double-helix resonance. At the same time, Eq. (51) corresponds to the (–) double-helix resonance due to the $\psi^{(1)} \leftrightarrow \psi^{(4)}$ interference. In this equation, one has $\nu^{(-)} = 1, 2, 3,...$ and $\beta^{(-)} \equiv \beta^{(1)} = \beta^{(4)}$. No solutions in a cylindrical coordinate system are presumes for the (–) double-helix resonance, $\psi^{(1)} \leftrightarrow \psi^{(4)}$, and a proper analysis can be made only in a helical coordinate system. For a bias magnetic field directed contrarily to $z$ axis, one obtains the $L$-MDM spectral characteristics based on solution of a system of two equations: Eq. (50) and Eq. (57) with $\beta^{(+)} \equiv \beta^{(2)} = \beta^{(3)}$. Since $\beta^{(2)} = \beta^{(3)} \neq \beta^{(1)} = \beta^{(4)}$, one can suppose that the absorption peak positions for the (+) resonances ($L$-MDMs) are different for the cases of a bias magnetic field directed along and contrarily to $z$ axis. This fact should give an evidence for the symmetry breakings for MDM oscillations in a quasi-2D ferrite disk. However, the question about experimental verification of the shift of the $L$-MDM absorption peaks for reversed directions of a bias magnetic field (along or contrarily to $z$ axis) is still open. A thickness $d$ of a real ferrite disk is very small compared to both $d_{eff}^{(1\leftrightarrow 4)}$ and $d_{eff}^{(2\leftrightarrow 3)}$. It can be presumed also that there is a very small difference between wavenumbers $\beta^{(1)} = \beta^{(4)}$ and $\beta^{(2)} = \beta^{(3)}$. All this may give negligibly small the bias-direction shift of the absorption spectra and very precise experiments should be made to verify this effect. To the best of our knowledge, no such experiments were realized till now.

There is, however, an experimental evidence for symmetry breakings of MDM oscillations for a given direction of a bias magnetic field following from experiments made about 50 years ago by Dillon [17]. In his experiments, Dillon used very well polished YIG-monocrystall disk. When such a normally magnetized disk was placed in a cavity in a region of a homogeneous microwave magnetic field, a multiresonance sharp-peak absorption spectrum was observed [17]. A similar experiment with a well polished YIG-monocrystall disk was repeated by Yukakawa and Abe [18]. One can suppose that these experimental spectra, both in Ref. [17] and in Ref. [18], are the spectra of the (+) resonant states: In these resonant states, magnetization precession coincides with direction of the field azimuth rotation and so such spectra should be more easily excited. When, however, Dillon used a well polished ferrite disk, but with a small radial flaw, the spectral peaks became split. This peak splitting is more evident for high-order modes. Similar peak splitting was observed in experimental spectra in Ref. [10]. In experiments [10], the authors used disks cut from an epitaxy-grown YIG-film wafer. Lateral surfaces of these disks were not specially polished and so had some imperfections. The peak splitting experimentally observed in Refs. [10, 17, 18], can be explained as follows. Due to a small radial flaw or small imperfections on a lateral surface, some transformation of one azimuthally rotating MDM to another one can occur in a ferrite disk. As a result, one has coupling between two double-helix-resonance modes, $\psi^{(1)} \leftrightarrow \psi^{(4)}$ and $\psi^{(2)} \leftrightarrow \psi^{(3)}$, which gives evident peak splitting. The fact of the presence of two double-helix resonances, $\psi^{(1)} \leftrightarrow \psi^{(4)}$ and $\psi^{(2)} \leftrightarrow \psi^{(3)}$, for a given direction of a bias magnetic field is an indirect evidence for symmetry breakings of MDM oscillations.

## 6. Conclusion

Long-range magnetic-dipolar interactions in confined magnetic structures are not in a scope of classical electromagnetic problems and, at the same time, have properties essentially different from the effects of exchange ferromagnetism. The MDM spectral properties in confined magnetic structures are based on postulates about a physical meaning of the MS-potential function $\psi(\vec{r},t)$ as a complex scalar wave function, which presumes the long-range phase correlations.

An important feature of the MDM oscillations in a ferrite disk concerns the fact that a structure with symmetric parameters and symmetric basic equations goes into eigenstates that are not space-time symmetric. In an analysis of spectral properties of MS-potential wave functions in quasi-2D ferrite disks one becomes faced with nonintegrability (path dependence) of the problem. This nonintegrability appears because of a special phase factor on a lateral border of a normally magnetized ferrite disk. One has a system without rotational and translational invariance and so without possibility for separation of variables in the wave equation. To make the MDM spectral problem analytically integrable, two approaches were suggested. These approaches, distinguishing by differential operators and boundary conditions for the MS-potential wave function, give two types of the MDM oscillation spectra. Solutions satisfying the second-order differential equation are conventionally called as *G*-mode magnetic-dipolar oscillations. These MDMs are characterized by the orthogonality relations with energy eigenstates and topological magnetic currents. Solutions satisfying the first-order differential-matrix operator we conventionally call as *L*-mode magnetic-dipolar oscillations. Proper integrable solutions for *L*-mode MS-potential functions can be obtained from an analysis of the MDM propagation in a helical coordinate system. This analysis discovers unique symmetry properties of MDM oscillations.

For helical waves, one has a combined effect of longitudinal and azimuth power flows. In our studies of MS helical waves, we use Waldron's helical coordinate system. In the Waldron coordinate system, the pitch of the helix is fixed but the pitch angle is allowed to vary as a function of the radius. Contrary to a cylindrical coordinate system, the helical coordinate system is not orthogonal. Inside a ferrite region, the MS-potential wave function is described by the Walker equation in helical coordinates. Outside a ferrite, one has the Laplace equation in helical coordinates. For a given quantity of a pitch, one obtains solutions for four MS helical waves. An analysis shows that there can be two pairs of reciprocally correlated (entangled) helical waves. For every of these pairs one can consider the closed-loop phase run of the double-helix resonance. In a disk, the quantity of pitch is determined by the virtual "reflection" planes, where the nodes or bulges of helical MS-potential wave functions can be located. For the entangled (double-helix) states there should be phase over-running for the azimuth coordinate, but no phase over-running for the MS-potential wave function along the disk axis. As a result, the helical-wave problem can be reduced to an integrable problem for effective membrane functions in a cylindrical coordinate system. The solutions give the MDM power-flow-density vortices with cores at the disk center and azimuthally running waves of magnetization.

The spectral properties of MDM oscillations analyzed in the paper give evidence for symmetry breaking effects. The solutions for MS-potential wave functions show non-time-invariant chiral states of magnetic structures. The theory presented in the paper clearly explains the experimentally observed splitting of the spectral peaks as a result of coupling between two double-helix-resonance modes. The presence of such splitting is an indirect evidence for symmetry breakings of MDM oscillations.

**Acknowledgement**

**References**

[1] V. G. Bar'yakhtar, V. A. L'vov, and D. A. Yablonskii, Pis'ma Zh. Eksp. Teor. Fiz. **37**, 565 (1983).
[2] A. M. Kadomtseva, A. K. Zvezdin, Yu. F. Popov1, A. P. Pyatakov1, and G. P. Vorob'ev, JETP Lett. **79**, 571 (2004).
[3] M. Mostovoy, Phys. Rev. Lett. **96**, 067601 (2006).
[4] *Electromagnetic, magnetostatic, and exchange-interaction vortices in confined magnetic structures*, Edited by E.O. Kamenetskii (Research Signpost Publisher, Kerala, India, 2008).
[5] E.O. Kamenetskii, Phys. Rev. E, **63**, 066612 (2001).
[6] E.O. Kamenetskii, M. Sigalov, and R. Shavit, J. Phys.: Condens. Matter **17**, 2211 (2005).
[7] E.O. Kamenetskii, J. Phys. A: Math. Theor. **40**, 6539 (2007).
[8] M. Sigalov, E.O. Kamenetskii, and R. Shavit, J. Phys.: Condens. Matter **21**, 016003 (2009).
[9] E.O. Kamenetskii, M. Sigalov, and R. Shavit, J. Appl, Phys. **105**, 013537 (2009).
[10] E. O. Kamenetskii, A.K. Saha, and I. Awai, Phys. Lett. A **332**, 303 (2004).
[11] M. Sigalov, E.O. Kamenetskii, and R. Shavit, Appl. Phys. B **93**, 339 (2008).
[12] M. Sigalov, E. O. Kamenetskii, and R. Shavit, J. Appl. Phys. **104**, 053901 (2008).
[13] A. Gurevich and G. Melkov, *Magnetic Oscillations and Waves* (CRC Press, New York, 1996).
[14] T. Morimae, A. Sugita, and A. Shimizu, Phys. Rev. A **71**, 032317 (2005).
[15] E. Sjöqvist, Phys. Rev. A **62**, 022109 (2000).
[16] B. Basu and P. Bandyopadhyay, J. Phys. A: Math. Theor. **41**, 055301 (2008).
[17] J. F. Dillon Jr., J. Appl. Phys. **31**, 1605 (1960).
[18] T. Yukawa and K. Abe, J. Appl. Phys. **45**, 3146 (1974).
[19] K. Rivkin, A, Heifetz, P. R. Sievert, and J. B. Ketterson, Phys. Rev. B **70**, 184410 (2004).
[20] H. Puszkarski, M. Krawczyk, and J.-C. S. Levy, Phys. Rev. B **71**, 014421 (2005); M. Krawczyk and H. Puszkarski, J. Appl. Phys. **104**, 113920 (2008).
[21] M. A. Naimark, *Linear Differential Operators*, Vol. 1 (Frederick Unger Publ. Co., New York, 1968); S. G. Mikhlin, *Variational Methods in Mathematical Physics* (McMillan, New York, 1964).
[22] J.D. Jackson, *Classical Electrodynamics*, 2nd ed. (Wiley, New York, 1975).
[23] P.H. Bryant, J. F. Smyth, S. Schultz, and D. R. Fredkin, Phys. Rev. B **47**, 11255 (1993).
[24] K.Yu. Guslienko, S. O. Demokritov and B. Hillebrands, and A. N. Slavin, Phys. Rev. B **66**, 132402 (2002).
[25] C.E. Zaspel, B. A. Ivanov, J. P. Park and P. A. Crowell, Phys. Rev. B **72**, 024427 (2005).
[26] K. Yu. Guslienko and A. H. Slavin, Phys. Rev. B **72**, 014463 (2005).
[27] G. T. Rado and J. R. Weertmann, J. Chem. Phys. Solids **11**, 315 (1959).
[28] J. Jeener, Phys. Rev. Lett. **82**, 1772 (1999).

[29] K. L. Sauer, F. Marion, P. J. Nacher, and G. Tastevin, Phys. Rev. B **63**, 184427 (2001).
[30] R. L. White and I. H. Solt, Jr., Phys. Rev. **104**, 56 (1956).
[31] L. R. Walker, Phys. Rev. **105**, 390 (1957).
[32] D.D. Osheroff and M.C. Cross, Phys. Rev. Lett. **59**, 94 (1987).
[33] L.D. Landau and E.M. Lifshitz, *Electrodynamics of Continuous Media*, 2nd ed. (Pergamon, Oxford, 1984).
[34] J. M. Luttinger and L. Tisza, Phys. Rev. **70**, 954 (1946).
[35] D. I. Kamenev, G. P. Berman, and V. I. Tsifrinovich, Phys. Rev. A **73**, 062336 (2006).
[36] R. A. Bertlmann, K. Durstberger, Y. Hasegawa, and B. C. Hiesmayr, Phys. Rev. A **69**, 032112 (2004).
[37] R.A. Waldron, Quarterly J. Mech. Appl. Math. **11**, 438 (1958).
[38] P. J. Lin-Chung and A. K. Rajagopal, Phys. Rev. E **52**, 901 (1995).
[39] P.L. Overfelt, Phys. Rev. E **64**, 036603 (2001).
[40] E. O. Kamenetskii, J. Magn. Magn. Mater. **302**, 137 (2006).
[41] E. O. Kamenetskii, arXiv:cond-mat/0505717.
[42] L. R. Walker, Phys. Rev. **105**, 390 (1957).
[43] J. F. Nye and M. V. Berry, Proc. R. Soc. London Ser. A **336**, 165 (1974).
[44] D. Belitz, T. R. Kirpatrick, and A. Rosch, Phys. Rev. B 73, 054431 (2006).
[45] E. O. Kamenetskii, J. Appl. Phys. **105**, 093913 (2009).

**Figure captions:**

Fig. 1. The right-handed (a) and left-handed (b) Waldron's helical coordinate systems. Coordinate $\zeta$ is measured parallel to coordinate $z$ from the reference surface $z = p\theta/2\pi$. Arrows illustrate, as an example, the propagation directions of helical waves: (a) the forward right-hand-helix MS wave (the wave $\psi^{(1)}$) and (b) the forward left-hand-helix MS wave (the wave $\psi^{(2)}$).

Fig. 2. The (+) resonance caused by the $\psi^{(1)} \leftrightarrow \psi^{(4)}$ phase correlation for $d_{eff}^{(1\leftrightarrow 4)}$ equal to $p^{(+)}/2$. The MS potential $\psi = \psi^{(1)} + \psi^{(4)} = 0$ in points *a* and *b*. The azimuth phase over-running for the EMF is characterized by the azimuth wave number $\nu^{(+)} = 1$. Arrows show directions of propagation for helical MS modes and a direction of rotation of a composition of helices. When a composition of waves $\psi^{(1)} \leftrightarrow \psi^{(4)}$ acquires a geometrical phase over-running of $2\pi$ during a period, every separate wave, $\psi^{(1)}$ or $\psi^{(4)}$, has a dynamical phase over-running of $\pi$ and so behaves as a double-valued function inside a disk.

Fig. 3. The (+) resonance caused by the $\psi^{(1)} \leftrightarrow \psi^{(4)}$ phase correlation for $d_{eff}^{(1\leftrightarrow 4)}$ equal to $p^{(+)}$. The MS potential $\psi = \psi^{(1)} + \psi^{(4)} = 0$ in points *a*, *b*, and *c*. The azimuth phase over-running for the EMF

is characterized by the azimuth wave number $v^{(+)}=1$. Arrows show directions of propagation for helical MS modes and a direction of rotation of a composition of helices.

Fig. 4. The (−) resonance caused by the $\psi^{(2)} \leftrightarrow \psi^{(3)}$ phase correlation for $d_{eff}^{(2\leftrightarrow 3)}$ equal to $p^{(-)}/2$. The MS potential $\psi = \psi^{(2)} + \psi^{(3)} = 0$ in points $a$ and $b$. Arrows show directions of propagation for helical MS modes and a direction of rotation of a composition of helices.

Fig. 5. Illustration of the (+) resonance in a real thin-film disk for the case when $d_{eff}^{(1\leftrightarrow 4)} = p^{(+)}/2$ and $v^{(+)}=1$. A real ferrite disk is an open thin-film structure with $\beta^{(+)}d \ll \beta^{(+)}d_{eff}^{(1\leftrightarrow 4)}$. The virtual "reflection" planes for helical modes are found in free space regions above and below a disk.

Fig. 6. Illustration of the (+) resonance in a real thin-film disk for the case when $d_{eff}^{(1\leftrightarrow 4)} = p^{(+)}$ and $v^{(+)}=1$. A real ferrite disk is an open thin-film structure with $\beta^{(+)}d \ll \beta^{(+)}d_{eff}^{(1\leftrightarrow 4)}$. The virtual "reflection" planes for helical modes are found in free space regions above and below a disk.

Fig. 7. The normalized "thickness" functions $\xi(z)$ for the 1$^{st}$ ($q = 1$) and 2$^{nd}$ ($q = 2$) MDMs.

Fig. 8. The effective membrane functions $\tilde{\varphi}$ for the 1$^{st}$ ($q = 1$) MDM at different time phases.

Fig. 9. The power flow density distribution for the 1$^{st}$ ($q = 1$) $L$-MDM calculated based on Eq. (69).

Fig. 10. Gallery of magnetization $\vec{m}$ at different time phases for the 1$^{st}$ ($q = 1$) MDM (arbitrary units).

Fig. 11. Gallery of magnetization $\vec{m}$ at different time phases for the 2$^{nd}$ ($q = 2$) MDM (arbitrary units).

Fig. 12. Explicit illustration of cyclic evolution of magnetization for the 1$^{st}$ ($q = 1$) MDM. When (for a given radius inside a disk and a given time phase $\omega t$) an azimuth angle $\theta$ varies from 0 to $2\pi$, the magnetization vector accomplishes the $2\pi$ geometric-phase rotation. For a given azimuth angle $\theta$, the magnetization vectors are in phase in the radial direction.

Fig. 13. Explicit illustration of cyclic evolution of magnetization for the 2$^{nd}$ ($q = 2$) MDM. When (for a given radius inside a disk and a given time phase $\omega t$) an azimuth angle $\theta$ varies from 0 to $2\pi$, the magnetization vector accomplishes the $2\pi$ geometric-phase rotation. For a given azimuth angle $\theta$, the magnetization vectors are in phase or $180°$ out of phase in the radial direction.

"Space-time symmetry violation of the fields in quasi-2D ferrite particles with magnetic-dipolar-mode oscillations", by E.O. Kamenetskii

-------------------------------------------------------------------------------------------------------------------

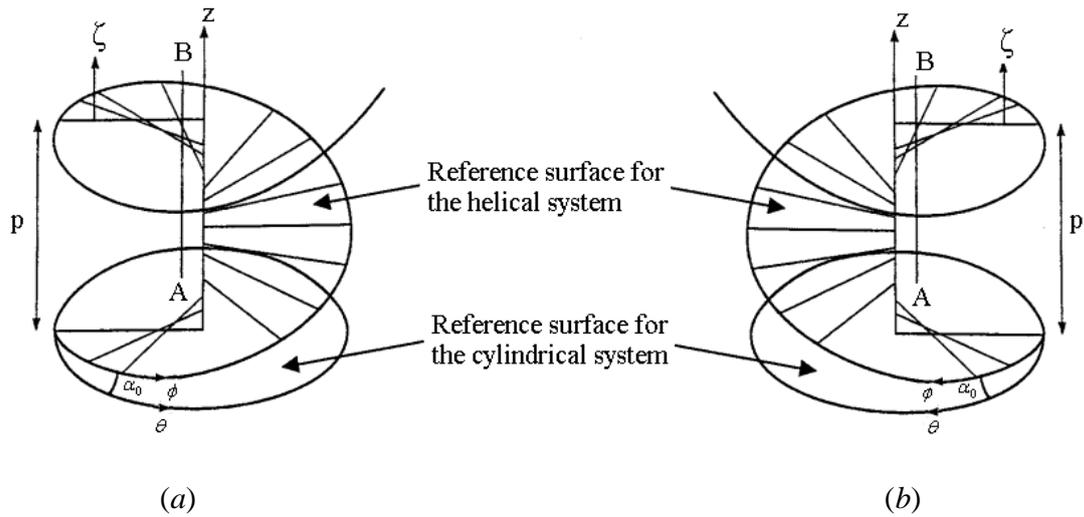

(a)                           (b)

Fig. 1. The right-handed (a) and left-handed (b) Waldron's helical coordinate systems. Coordinate $\zeta$ is measured parallel to coordinate $z$ from the reference surface $z = p\theta/2\pi$. Arrows illustrate, as an example, the propagation directions of helical waves: (a) the forward right-hand-helix MS wave (the wave $\psi^{(1)}$) and (b) the forward left-hand-helix MS wave (the wave $\psi^{(2)}$).

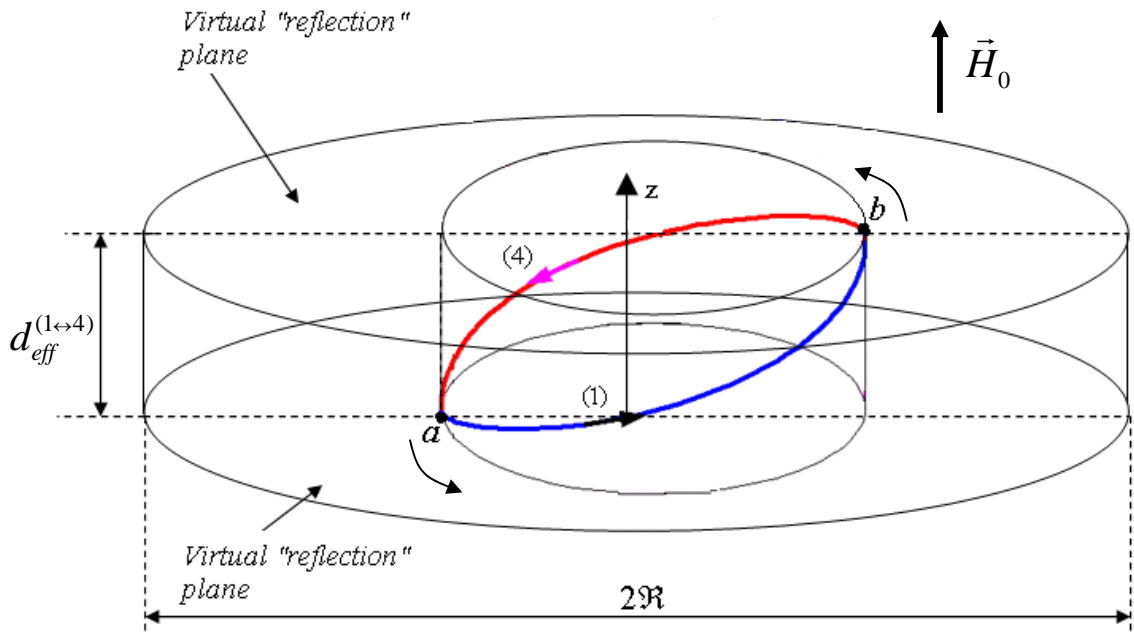

Fig. 2. The (+) resonance caused by the $\psi^{(1)} \leftrightarrow \psi^{(4)}$ phase correlation for $d_{eff}^{(1 \leftrightarrow 4)}$ equal to $p^{(+)}/2$. The MS potential $\psi = \psi^{(1)} + \psi^{(4)} = 0$ in points $a$ and $b$. The azimuth phase over-running for the EMF is characterized by the azimuth wave number $\nu^{(+)} = 1$. Arrows show directions of propagation for helical MS modes and a direction of rotation of a composition of helices. When a composition of waves $\psi^{(1)} \leftrightarrow \psi^{(4)}$ acquires a geometrical phase over-running of $2\pi$ during a period, every separate wave, $\psi^{(1)}$ or $\psi^{(4)}$, has a dynamical phase over-running of $\pi$ and so behaves as a double-valued function inside a disk.

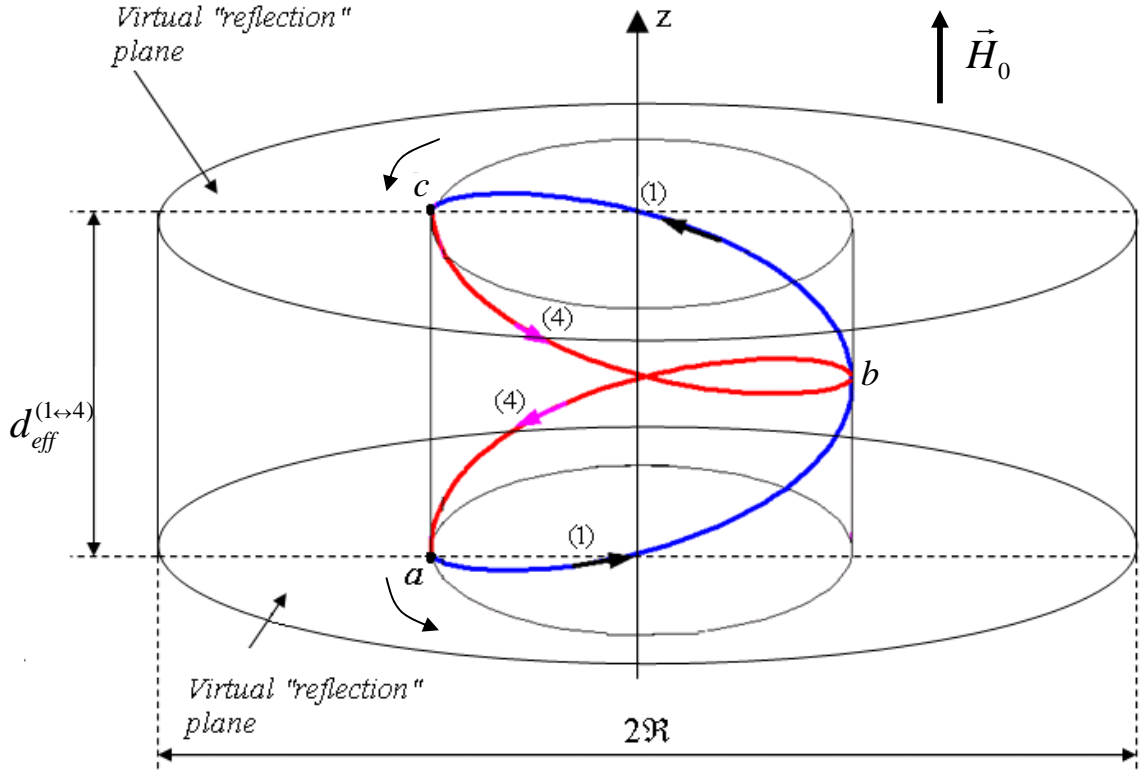

Fig. 3. The (+) resonance caused by the $\psi^{(1)} \leftrightarrow \psi^{(4)}$ phase correlation for $d_{eff}^{(1\leftrightarrow 4)}$ equal to $p^{(+)}$. The MS potential $\psi = \psi^{(1)} + \psi^{(4)} = 0$ in points $a$, $b$, and $c$. The azimuth phase over-running for the EMF is characterized by the azimuth wave number $\nu^{(+)} = 1$. Arrows show directions of propagation for helical MS modes and a direction of rotation of a composition of helices.

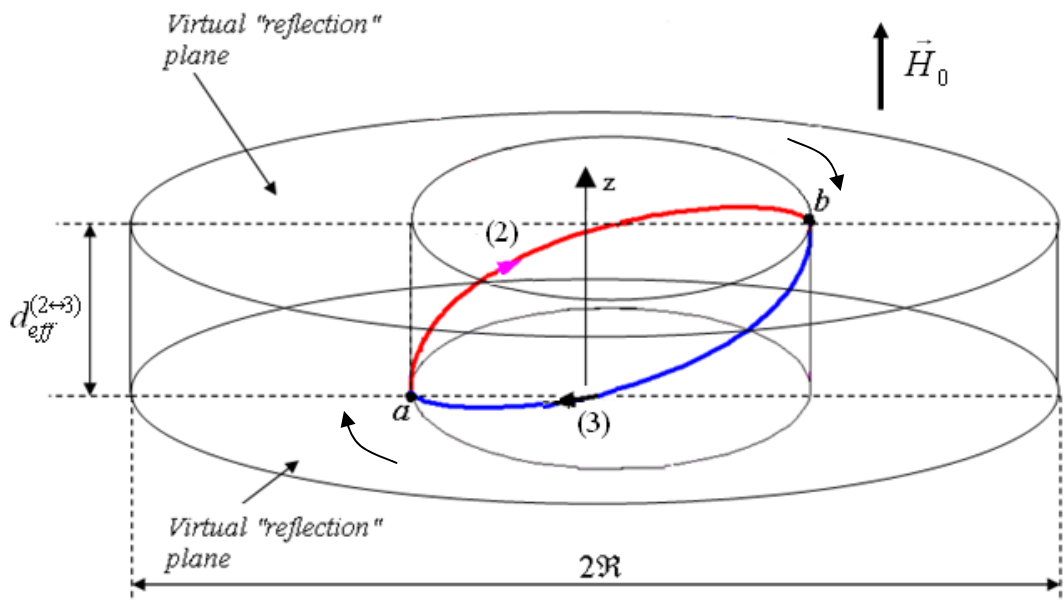

Fig. 4. The (–) resonance caused by the $\psi^{(2)} \leftrightarrow \psi^{(3)}$ phase correlation for $d_{eff}^{(2\leftrightarrow 3)}$ equal to $p^{(-)}/2$. The MS potential $\psi = \psi^{(2)} + \psi^{(3)} = 0$ in points *a* and *b*. Arrows show directions of propagation for helical MS modes and a direction of rotation of a composition of helices.

"Space-time symmetry violation of the fields in quasi-2D ferrite particles with magnetic-dipolar-mode oscillations", by E.O. Kamenetskii

-----------------------------------------------------------------------------------------------------------------------

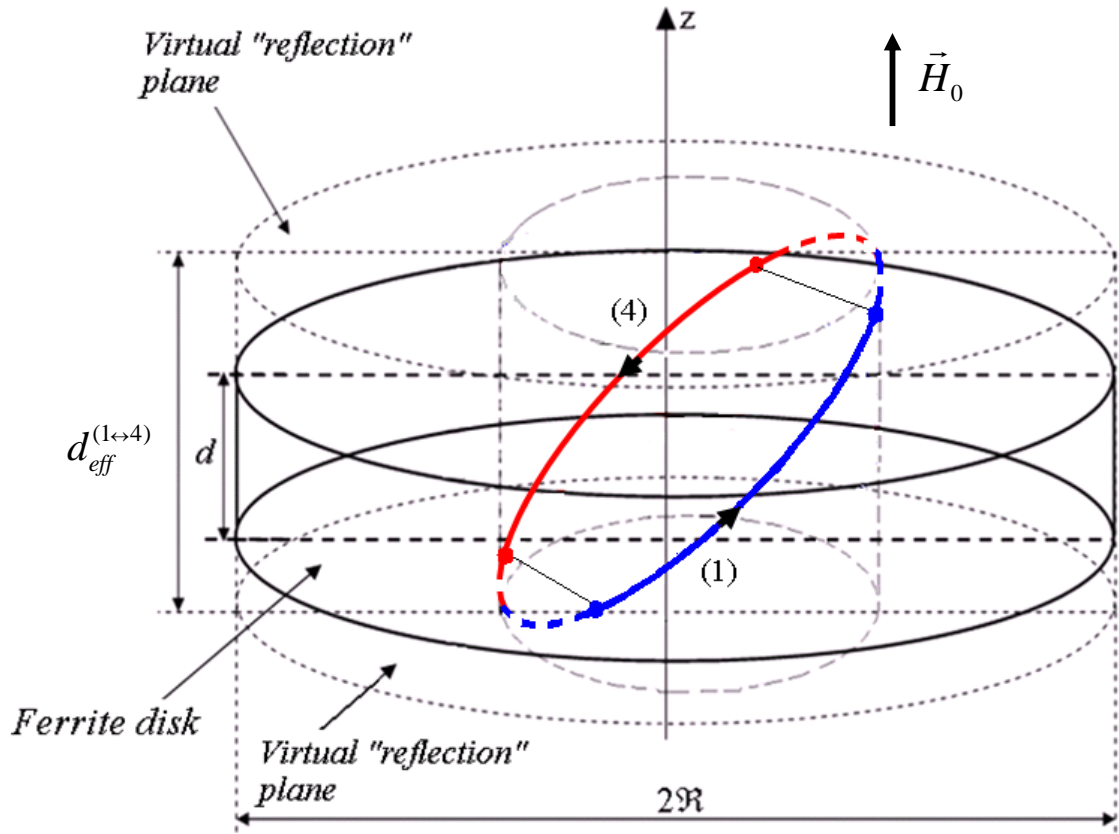

Fig. 5. Illustration of the (+) resonance in a real thin-film disk for the case when $d_{eff}^{(1\leftrightarrow 4)} = p^{(+)}/2$ and $\nu^{(+)} = 1$. A real ferrite disk is an open thin-film structure with $\beta^{(+)}d << \beta^{(+)}d_{eff}^{(1\leftrightarrow 4)}$. The virtual "reflection" planes for helical modes are found in free space regions above and below a disk.

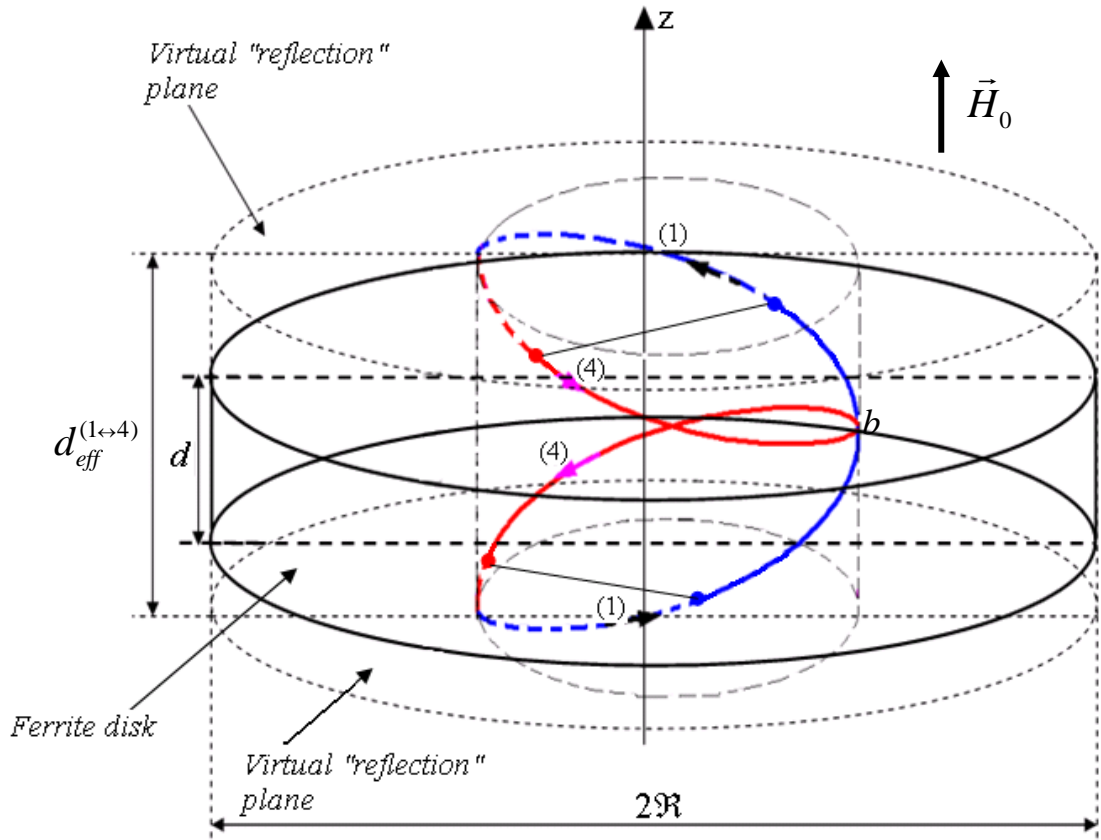

Fig. 6. Illustration of the (+) resonance in a real thin-film disk for the case when $d_{eff}^{(1\leftrightarrow 4)} = p^{(+)}$ and $v^{(+)} = 1$. A real ferrite disk is an open thin-film structure with $\beta^{(+)}d << \beta^{(+)}d_{eff}^{(1\leftrightarrow 4)}$. The virtual "reflection" planes for helical modes are found in free space regions above and below a disk.

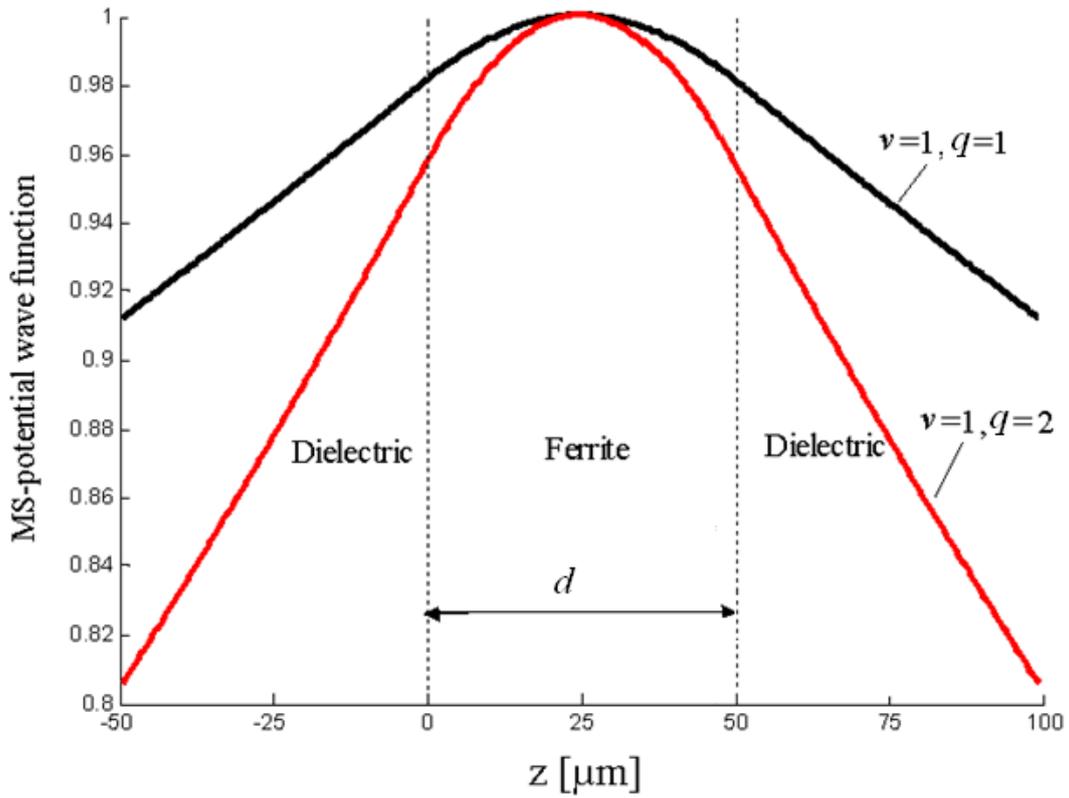

Fig. 7. The normalized "thickness" functions $\xi(z)$ for the 1$^{st}$ ($q = 1$) and 2$^{nd}$ ($q = 2$) MDMs.

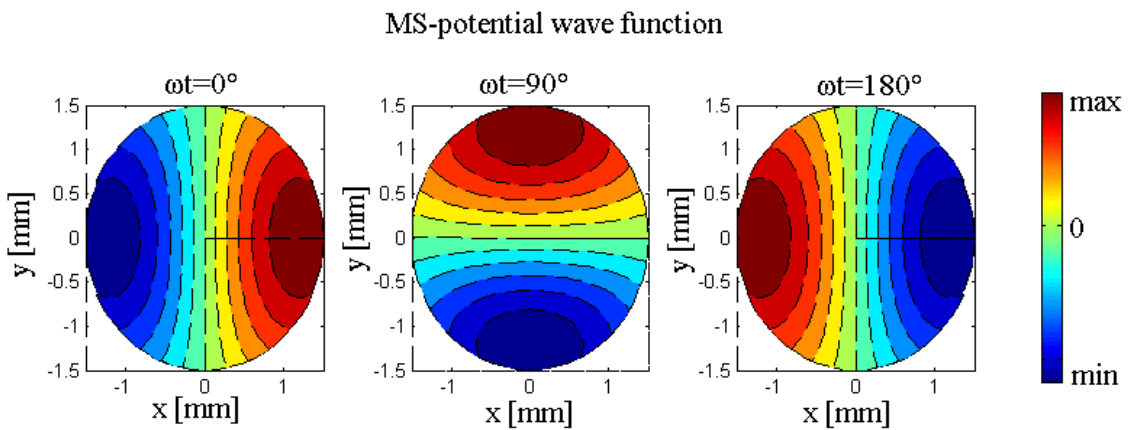

Fig. 8. The effective membrane functions $\tilde{\varphi}$ for the 1$^{st}$ ($q = 1$) MDM at different time phases.

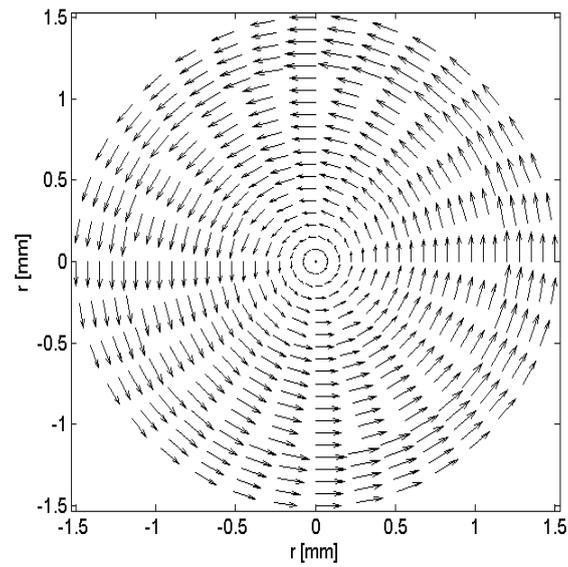

Fig. 9. The power flow density distribution for the 1$^{st}$ ($q = 1$) $L$-MDM calculated based on Eq. (69).

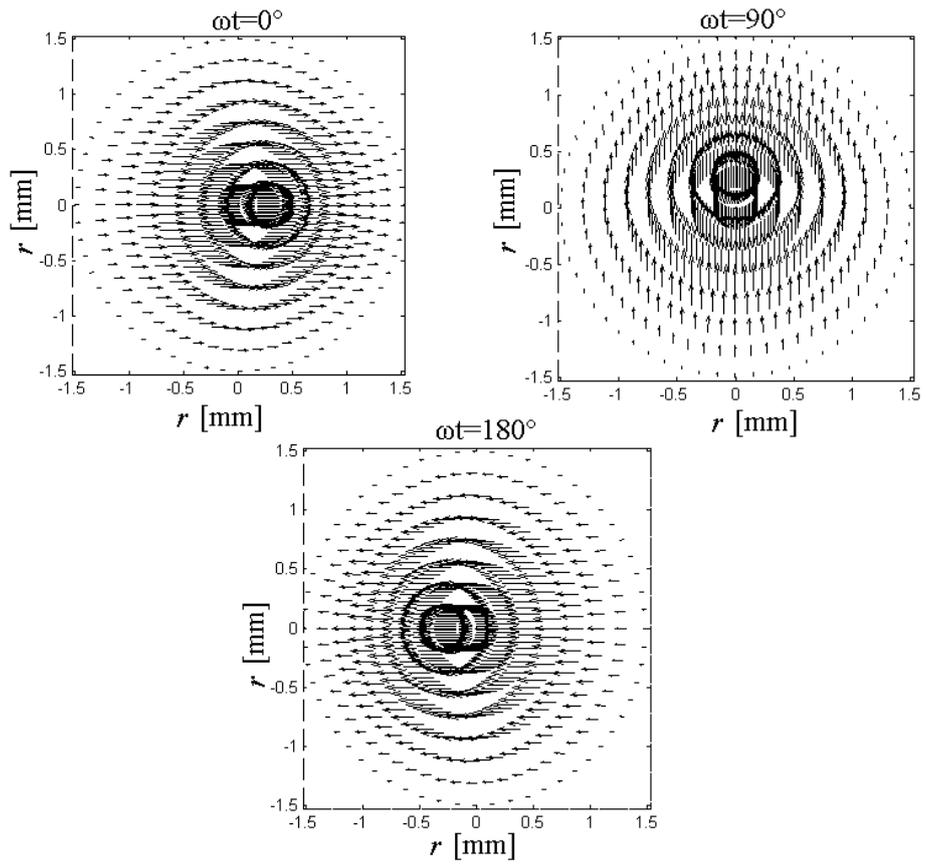

Fig. 10. Gallery of magnetization $\vec{m}$ at different time phases for the 1$^{\text{st}}$ ($q = 1$) MDM.

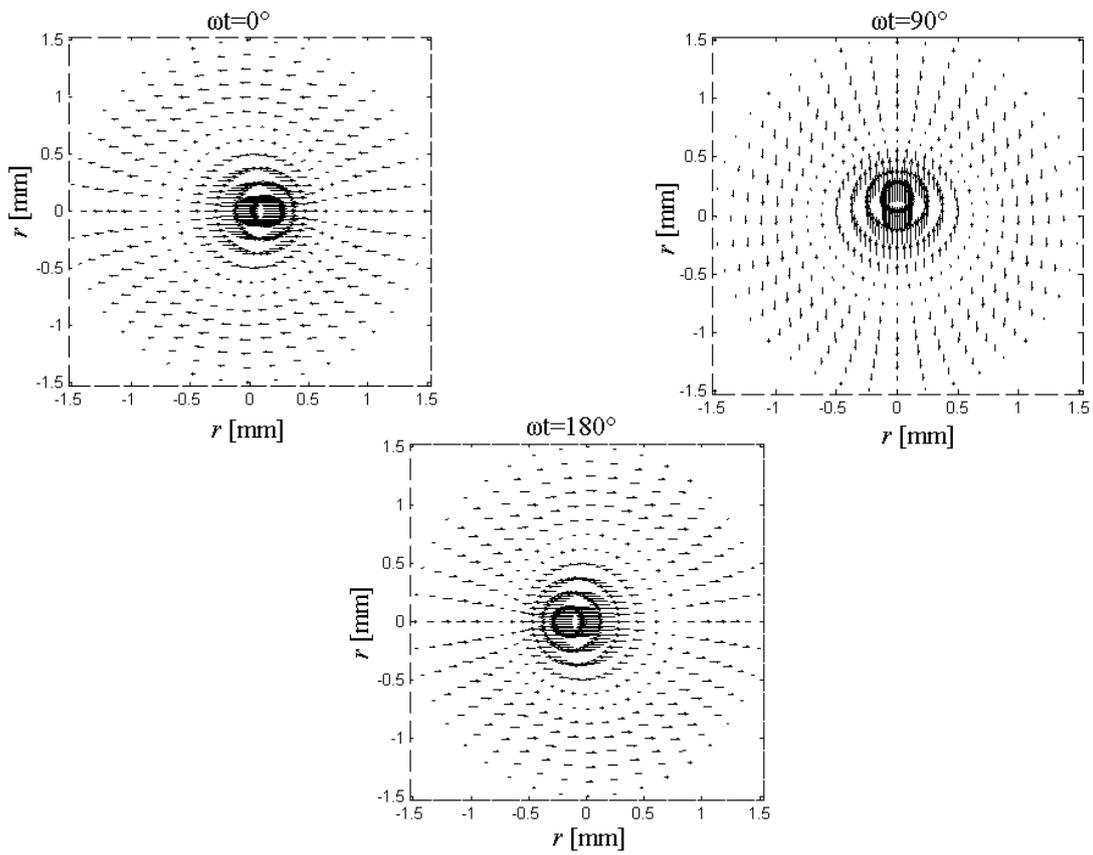

Fig. 11. Gallery of magnetization $\vec{m}$ at different time phases for the $2^{nd}$ ($q = 2$) MDM.

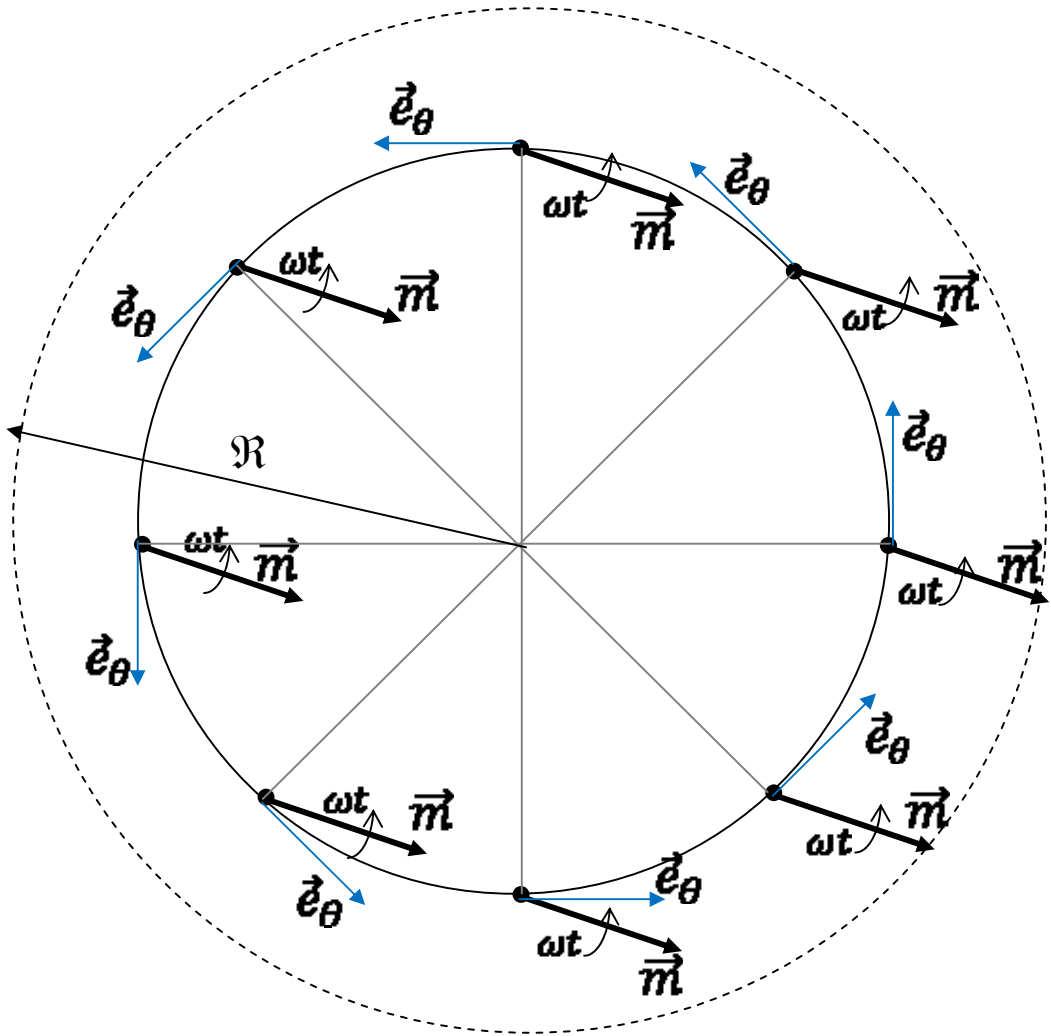

Fig. 12. Explicit illustration of cyclic evolution of magnetization for the 1$^{st}$ ($q = 1$) MDM. When (for a given radius inside a disk and a given time phase $\omega t$) an azimuth angle $\theta$ varies from 0 to $2\pi$, the magnetization vector accomplishes the $2\pi$ geometric-phase rotation. For a given azimuth angle $\theta$, the magnetization vectors are in phase in the radial direction.

"Space-time symmetry violation of the fields in quasi-2D ferrite particles with magnetic-dipolar-mode oscillations", by E.O. Kamenetskii

---

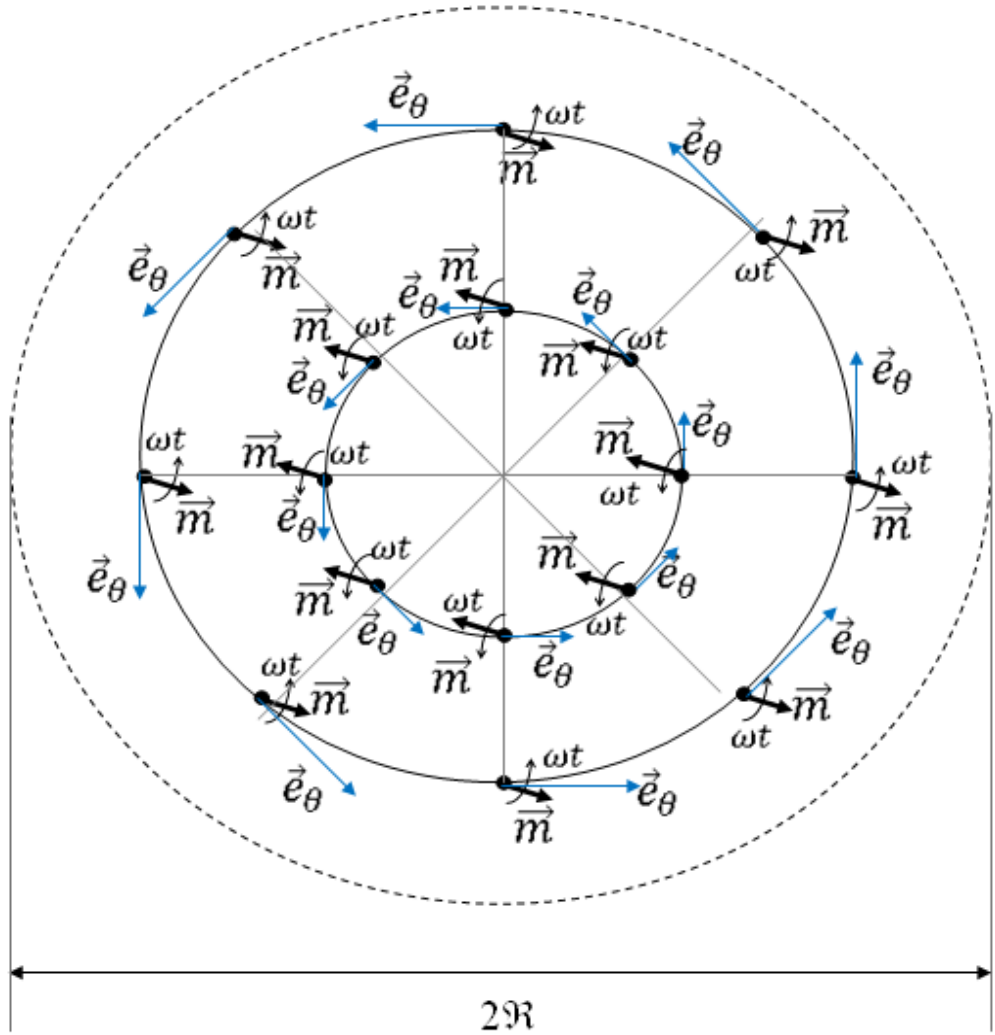

Fig. 13. Explicit illustration of cyclic evolution of magnetization for the 2$^{nd}$ ($q = 2$) MDM. When (for a given radius inside a disk and a given time phase $\omega t$) an azimuth angle $\theta$ varies from 0 to $2\pi$, the magnetization vector accomplishes the $2\pi$ geometric-phase rotation. For a given azimuth angle $\theta$, the magnetization vectors are in phase or 180° out of phase in the radial direction.

40